\documentclass[useAMS,usenatbib]{mn2e}

\usepackage{times,epsfig,natbib,amssymb,amsmath,graphics,longtable,threeparttable}

\usepackage{rotating}

\def \lsun{\ifmmode{{\rm\ L}_\odot}\else{${\rm\ L}_\odot $}\fi}

\def \msun{\ifmmode{{\rm\ M}_\odot}\else{${\rm\ M}_\odot$}\fi}

\def \rsun{\ifmmode{{\rm\ R}_\odot}\else{${\rm\ R}_\odot$}\fi}

\newcommand{\kms}{kms$^{-1}$}                         

\def \mdot{\ifmmode{{\rm\dot{M}}}\else{${\rm\dot{M}}$}\fi}

\newcommand{\ha}{H$\alpha${}}

\newcommand{\nii}{[N\,{\sc ii}]}

\title[CC SN environments]{Progenitor mass constraints for core-collapse
supernovae from correlations with host galaxy star formation\thanks{Based 
on observations made with the Isaac Newton Telescope
operated on the island
of La Palma by the Isaac Newton Group in the Spanish Observatorio del Roque de los
Muchachos of the Instituto de Astrofisica de Canarias, observations made with
the Liverpool 
Telescope operated on the island of La Palma by Liverpool John Moores
University 
in the Spanish Observatorio del Roque de los Muchachos of the Instituto de 
Astrofisica de Canarias with financial support from the UK Science and
Technology Facilities Council, and observations made with the 2.2m
MPG/ESO telescope at La Silla, proposal ID: 084.D-0195.}}
\author[Anderson et al.]{J. P. Anderson$^{1}$\thanks{E-mail:
anderson@das.uchile.cl}, S. M. Habergham$^{2}$, P. A. James$^{2}$ \&\ M. Hamuy$^{1}$\\
$^{1}$Departamento de Astronom\'ia, Universidad de Chile, Casilla 36-D, 
Santiago, Chile\\
$^{2}$Astrophysics Research Institute,
Liverpool John Moores University,
Twelve Quays House,
Egerton Wharf,
Birkenhead,
CH41 1LD,
UK
}

\begin{document}

\date{}

\pagerange{\pageref{firstpage}--\pageref{lastpage}} \pubyear{2012}

\maketitle

\label{firstpage}

\begin{abstract}
Using \ha\ emission as a tracer of on-going ($<$16 Myr old) and near-UV emission as a tracer
of recent (16-100 Myr old) star formation, we present constraints on the properties of
core-collapse supernova progenitors through the association of their
explosion sites with star forming regions. Amalgamating previous
results with those gained from new data, we present statistics of a large sample of
supernovae; 163.5 type II (58 IIP, 13 IIL, 13.5 IIb, 19 IIn and 12
`impostors', plus 48 with no sub-type classification) and 96.5 type Ib/c (39.5
Ib and 52 Ic, plus 5 with no sub-type classification). Using pixel
statistics we build distributions of associations of different supernova
types with host galaxy star formation. Our main findings and conclusions are:\\
1) An increasing progenitor mass sequence is observed, implied from an
increasing association of supernovae to host galaxy \ha\ emission. This
commences with the type Ia showing the weakest association,
followed by the type II, then the Ib, with the Ic showing the strongest correlation
to star forming regions. Thus our progenitor mass sequence runs
Ia-II-Ib-Ic.\\
2) Overall the type Ibc supernovae are found to occur nearer to bright HII 
regions than supernovae of type II. This implies that the former have shorter
stellar lifetimes thus arising from more massive progenitor stars.\\
3) While type IIP supernovae do not closely follow the on-going star formation, 
they accurately trace the recent formation. This implies that their 
progenitors arise from stars at the low end of the CC SN mass sequence,
consistent with direct detections of progenitors in pre-explosion imaging.\\
4) Similarly the type IIn supernovae trace recent but not the on-going
star formation. This implies that, contrary to the general consensus, the
majority of these
supernovae do \textit{not} arise from the most massive stars.\\
Results and suggestive constraints are also presented for the less numerous supernovae of
types IIL, IIb,  and supernova `impostors'. Finally we present analysis of possible 
biases in the data, the
results of which argue strongly against any selection effects that could 
explain the relative excess of type Ibc supernovae within bright HII regions. 
Thus intrinsic progenitor differences in the sense of the mass sequence we propose 
remain the most plausible explanation of our findings.
\end{abstract}


\section{Introduction}
\label{intro}
Core-collapse (CC) supernovae (SNe) are the explosive fate of
massive ($>$8-10\msun) stars. This occurs after successive stages of nuclear
fusion lead to the formation of a degenerate iron core. Nuclear fusion is 
then energetically unfavorable and continued energy losses accelerate the
collapse of the core. The bounce of this collapse once nuclear
densities are reached, and its subsequent interaction
with still infalling material aided by copious neutrino fluxes, is
thought to produce a shock wave that expels
the star's envelope generating the transients we observe (see \citealt{mez05}, for a review
of the latter stages of stellar evolution that lead to core collapse and
SN)\footnote{Whether 
CC explosions have actually been observed
in models, for the full range of progenitor masses, is still under debate. The reader is
referred to the recent literature which discuss the explosion processes in
detail (e.g. \citealt{bru09,nor10,han11}).}.\\
The immense temperatures, densities and energies reached during the
explosive CC process lead to CC SNe playing a huge role in defining the
evolution of their environments, and hence the Universe. They are responsible
for a significant fraction of heavy elements produced,
and their energetics impact into their local environments, driving galaxy
evolution and possibly triggering further star formation (SF).\\ 
However, the community is still far
from agreement on which types of progenitors give rise to the
rich diversity in transient light curves and spectra we observe. Mapping
the paths between progenitor characteristics and transient
phenomena has become a key goal of SN science that is rapidly
increasing our understanding of stellar evolution and
the role parameters such as metallicity, rotation and binarity play in the
final stages of the lives of
massive stars.\\

\subsection{CC SN types}
\label{types}
SNe were initially
separated into types I and II through the absence or presence of hydrogen 
in their spectra \citep{min41}. While all type II events are thought to 
be produced through the CC mechanism, in addition the
types Ib and Ic of the non-hydrogen class are also believed to occur through CC.
Hence when speaking on the overall CC family we are discussing those SNe
classified as II, Ib or Ic. 
SNe types Ib and Ic (throughout the rest of the paper when referring to `SNIbc'
we are discussing the group of all those SNe that are classified in the
literature as `Ib', `Ic' or `Ib/c') lack strong silicon absorption seen in
SNe type Ia (SNIa; thermonuclear events), while the two are distinguished by the presence
(in the former) and absence (in the latter) of helium in their spectra
(see \citealt{fil97} for a review of SN spectral classifications). SNe
type II (SNII henceforth) can be further split into various
sub-types. SNIIP and SNIIL are differentiated by their
light-curve shapes, with the former showing a plateau and the latter a linear
decline \citep{bar79}. SNIIn show narrow emission features in their
spectra \citep{sch90}, indicative of interaction with pre-existing, 
slow-moving circumstellar material (CSM, \citealt{chu94}). SNIIb are
transitional objects as at early times they show hydrogen features, while
later 
this hydrogen disappears and hence their spectra
appear similar to SNIb \citep{fil93}. Finally,
there is
a group of objects known as `SN impostors'. These are transient objects which
when discovered appear to be similar to SNe (i.e. events that are the
end-points of a star's evolution), but are likely to be 
non-terminal, and hence possibly recurring,
explosions from massive stars (e.g. \citealt{van00,mau06}, although see \citealt{koc12}).\\
While the exact reasons for differences in light-curves and
spectra are unknown, they are likely to be the product
of differences in initial progenitor characteristics (mass,
metallicity, binarity and rotation) which affect the stellar evolution of
the star prior to SN. Effectively, these differences change the final structure of the star and
its surroundings prior to explosion and hence produce the diversity of
transients we observe. The main parameter affecting the nature and evolution
of the light we detect from a SN would appear to be the amount of stellar
envelope that is left at the epoch of explosion, with the SNIIP retaining the
most and the SNIc retaining the least. This outer envelope
can be lost through stellar winds, 
or through mass transfer in a binary system. These processes are then
dependent on 4 main progenitor characteristics: initial mass, metallicity,
stellar rotation and the presence/absence of a close binary companion. 
Given the likely strong correlation between pre-SN mass loss and resultant SN
type it has been argued (see e.g. \citealt{che06}) that the CC classification
scheme can be placed in a sequence of increasing pre-SN mass-loss such as
follows:\\

\begin{center}
\textbf{SNIIP $\rightarrow$ IIL $\rightarrow$ IIb $\rightarrow$ IIn $\rightarrow$ Ib $\rightarrow$ Ic}
\end{center}

Our understanding of the accuracy of this picture and how it relates to progenitor
mass or other characteristics is far from complete. However, it is a useful starting point
and we will refer back to this sequence when later discussing our results
which imply differences in progenitor lifetimes and hence initial
masses.

\subsection{CC SN progenitor studies}
\label{prevprog}
The most direct evidence on progenitor properties is gained from finding
pre-SN stars on pre-explosion images after a nearby SN is
discovered. This has had success in a number of cases (see
e.g. \citealt{eli11} and \citealt{mau11} for recent examples, and
\citealt{sma09b} for a review on the subject), and long-term we are likely to
gain the best insights into progenitor properties through this
avenue.
However, while the direct detection approach can give important information on 
single events, the need for very nearby
objects limits the statistics gained from these studies. 
Therefore it is useful to
explore other avenues 
to further our understanding of SN progenitors from a statistical
viewpoint.\\
An easy to measure parameter of a single SN is the host galaxy
within which it occurs. This separation of events by galaxy type
was one of the initial reasons for the separation into
CC (requiring a massive star) and thermonuclear (arising
from a WD system) events, through the absence of the former in ellipticals
(see e.g. \citealt{van05}) where the stellar population is almost exclusively dominated by 
evolved stars\footnote{A small number of CC SNe have been detected
in galaxies classified as ellipticals, i.e. non star-forming. However, in all of these
galaxies there is some evidence of recent SF
\citep{hak08}.}. More detailed studies have investigated how the relative 
SN rates change with, e.g. luminosities of host galaxies 
(e.g. \citealt{pra03,boi09,arc10}), to infer metallicity trends, or have obtained
`direct' host galaxy metallicity measurements through spectroscopic
observations \citep{pri08b}. These valuable studies allow the inclusion of large
numbers of SNe enabling statistically significant trends to be
observed. However, in a typical star forming galaxy there are many distinct stellar
populations, each with its own characteristic age, metallicity and possibly
binary fraction. Therefore,
to infer differences in progenitor properties a number of assumptions have to
be made. For CC SNe, which have sufficiently short delay-times (period
between epoch of SF and observed transient) that the
position at which an event is found is close to its birth site, it is
arguably more important to attempt to characterise the exact environment
\textit{within} host galaxies where SNe are found, in order to pursue progenitor studies.\\
This is the approach we proceed with in the current study; investigating
correlations of SN type with the characteristics of their environments within
host galaxies in order to constrain progenitor properties. 

\subsubsection{Core-collapse SN environments}
\label{CCenv}
The distribution of massive stars in galaxies is traced by the presence of HII
regions and OB associations. Therefore, one can analyse the distribution of SNe
within host galaxies and compare these with those of
massive stars in order to attempt to constrain the former's progenitors. After
initial studies by \cite{ric84} and \cite{hua87}, \cite{van92} was the first to 
attempt to separate
CC SNe into SNII and SNIbc with respect to host environments. He
found no statistical difference between the association of the two with 
HII regions, albeit with low
statistics. Further studies were achieved by \cite{bar94} and \cite{van96} who 
again concluded that
the degrees of association of the two types were similar and hence
suggested that both types arose from similar mass progenitors (later a
detailed discussion will be presented on how these associations can be
interpreted). These studies
measured distances to nearby HII regions to gauge associations with
massive star populations. While this approach can give useful
information on individual SNe associations, the irregular nature and large range of
intrinsic sizes/luminosities of HII regions mean that objectively applying
this technique can be difficult.\\
More recently, a pixel statistics technique
has been developed \citep{jam06,fru06} which allows one to investigate SN
distributions in a more systematic way. \cite{kel08} used this technique to
investigate the association of SNe and long-duration Gamma Ray Bursts (LGRBs) with host $g$-band
light. They found that while the SNIa, SNII and SNIb
followed the $g$-band distribution, the SNIc and LGRBs were found to
occur more frequently on peaks of the flux, and hence they concluded
that both type of events arose from similarly massive progenitors. 
\cite{ras08} compared these distributions with predicted characteristic
stellar ages 
from analytical models of star-forming galaxies, 
and derived mass limits for CC
SNe, concluding that SNIc arise from progenitors with masses higher than
25\msun. \cite{lel10} looked specifically at the distribution of SNIbc
locations and compared them with those of Wolf-Rayet (WR) stars
in nearby galaxies, claiming that SNIbc were consistent with being produced by these stars.\\
In addition to investigating correlations of SNe with certain stellar
populations, one can look at radial distributions of events and attempt to
infer progenitor properties, by comparing these to parameters such
as metallicity gradients within galaxies. Following \cite{bar92}, \cite{van97}
found a suggestion that SNIbc are more centrally concentrated within host
galaxies than SNII, a result which has been confirmed with increased statistics
by \cite{tsv04}, \cite{hak08}, \cite{and09} and \cite{boi09}. 
These differences have 
generally been ascribed to a metallicity dependence in producing additional
SNIbc at the expense of SNII in the centres of galaxies which are believed to
have
enhanced (compared to the outer disk regions) metal abundances (see \citealt{hen99}).
These trends are expected on
theoretical grounds as at higher metallicity progenitor stars have higher
mass-loss rates through radiatively driven winds (see
e.g. \citealt{pul96,kud00,mok07}) and hence it is easier for a star to lose
its outer envelopes and explode as a SNIbc. However, this simple interpretation has
been questioned by recent work separating galaxies by the presence
of interaction or disturbance \citep{hab10}. This analysis found that the
centralisation is much more apparent in disturbed galaxies. Given that these
disturbed galaxies are likely to have much shallower (if any) metallicity
gradients \citep{kew10} it was concluded that there was an IMF effect
at play. A detailed follow-up paper addressing these issues is being
prepared 
(Habergham et al. in preparation).\\
A more direct way to measure environment metallicities is to obtain spectra of
the immediate environments of SNe and derive gas-phase metallicities from
emission line diagnostics. This approach was first achieved by \cite{mod08}
who looked specifically at the environments of GRBs and the broad-line 
SNIc which are
associated with GRBs. Subsequent works have searched for
differences between the CC types. \cite{mod11_2} and \cite{lel11} 
investigated differences between SNIb and SNIc
but came to different conclusions on whether there was any clear metallicity
difference, while \cite{and10} included a sample of SNII in addition to SNIbc 
and found 
that the SNIbc 
show only a small, barely significant offset to higher metallicity 
than SNII, while also finding little
difference between the SNIb and SNIc\footnote{Combining all
the SNIbc from these studies and including only those where measurements are 
at the exact site of the SN, it has been shown that there is indeed a
difference in the metallicities, with the SNIc
usually found in regions of higher chemical abundance \citep{mod12}.}. These studies are continuing, with care being taken to
include SNe from all types of host galaxy in order to remove possible biases (see
\citealt{mod11_2}, for a recent review of these results).\\
Most recently \cite{kel11} have combined many of the above techniques 
by using images and spectra from the Sloan Digital
Sky Survey (SDSS) to investigate environmental colours and host galaxy
properties of SNe of different types, and have used these observations to infer differences in
progenitor properties.\\ 
Our first contribution to this growing field was
published in \cite{jam06}. Here we introduced a pixel statistic (which is the
main analysis tool used for the current investigation), 
and used this together with other analyses to
investigate how SNe are associated with SF within their host galaxies, as
traced by \ha\ emission. 
The CC SN pixel statistics analysis of this work was
then updated with an increased sample size of 160 CC SNe in \cite{and08} (AJ08
hereafter). 
In the current paper we repeat this analysis but with significantly larger
samples. The current analysis is achieved on a sample of 260 CC SNe. After a
thorough search of the literature this sample can be broken down to 58
SNIIP, 13 IIL, 13.5 IIb (one SN, 2010P is classified as IIb/Ib in the literature and
is therefore added at half weighting to both distributions), 19 IIn, 12 SN
`impostors' and 48 SNe that only have `II' classification, plus 39.5 SNIb, 52
Ic and 5 SNe which only have `Ib/c' classification in the literature. We
analyse the association of all these SNe to SF within their host galaxies and
use these results to infer properties of their respective progenitors.\\
The rest of the paper is arranged as follows. Next we summarise the data used
and reduction processes applied, followed by a summary of our pixel statistics
technique in section 3. Then we present our results in section 4, followed by
a discussion of their implications for the properties of CC SN progenitors in
section 5. Finally we list our conclusions in section 6.\\

\begin{figure*}
\includegraphics[width=5.7cm]{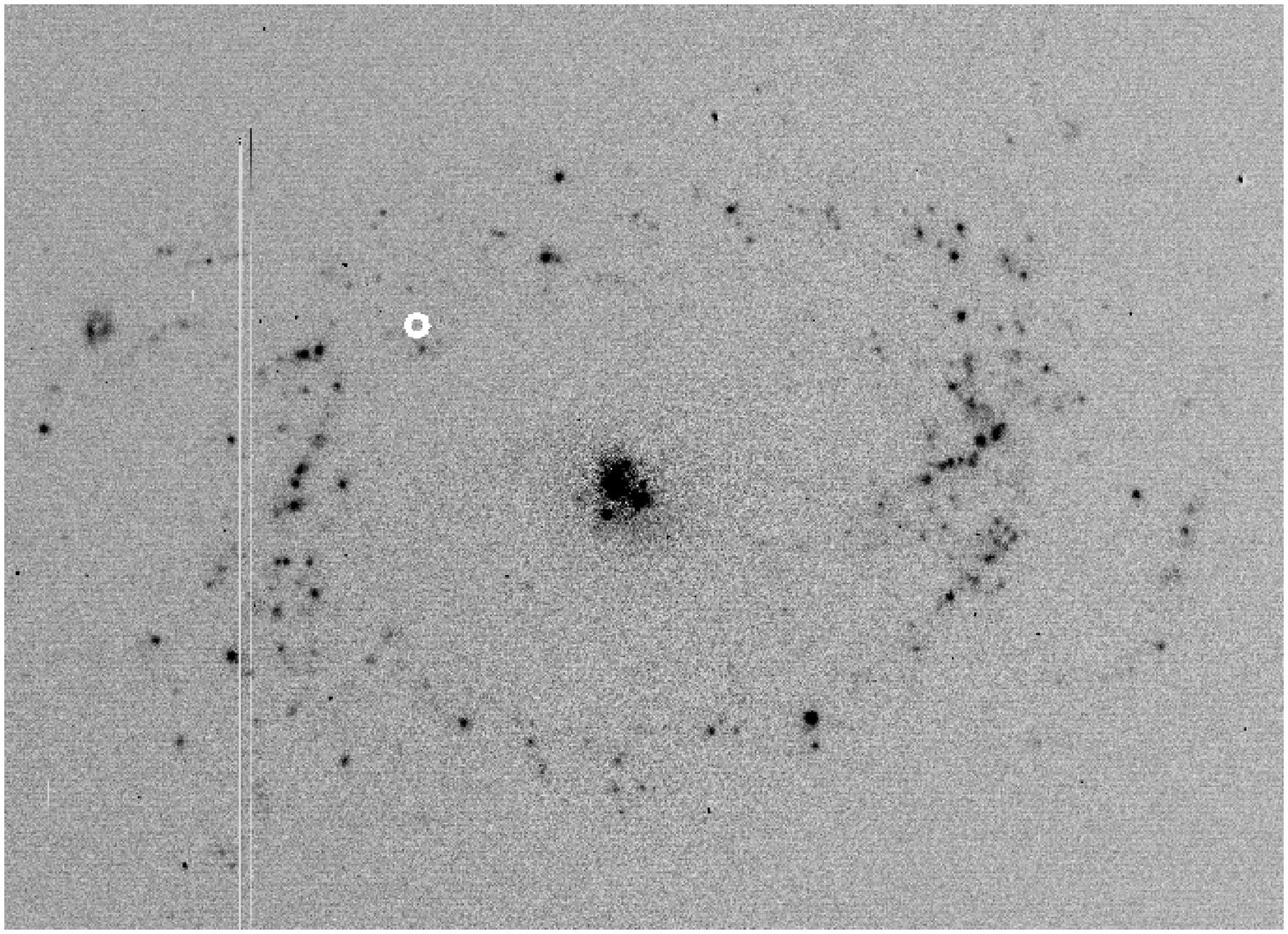}
\includegraphics[width=5cm]{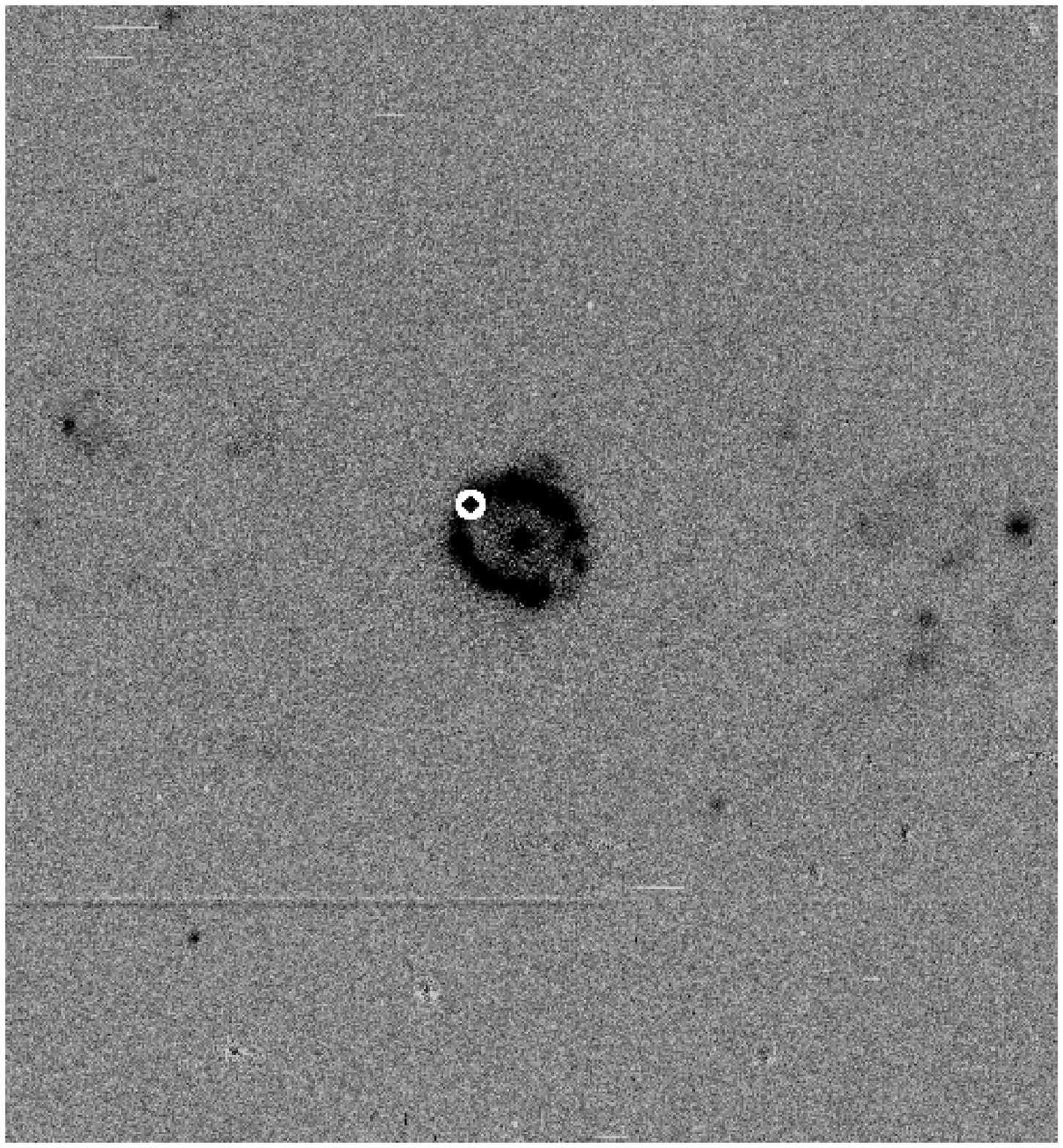}
\includegraphics[width=5.5cm]{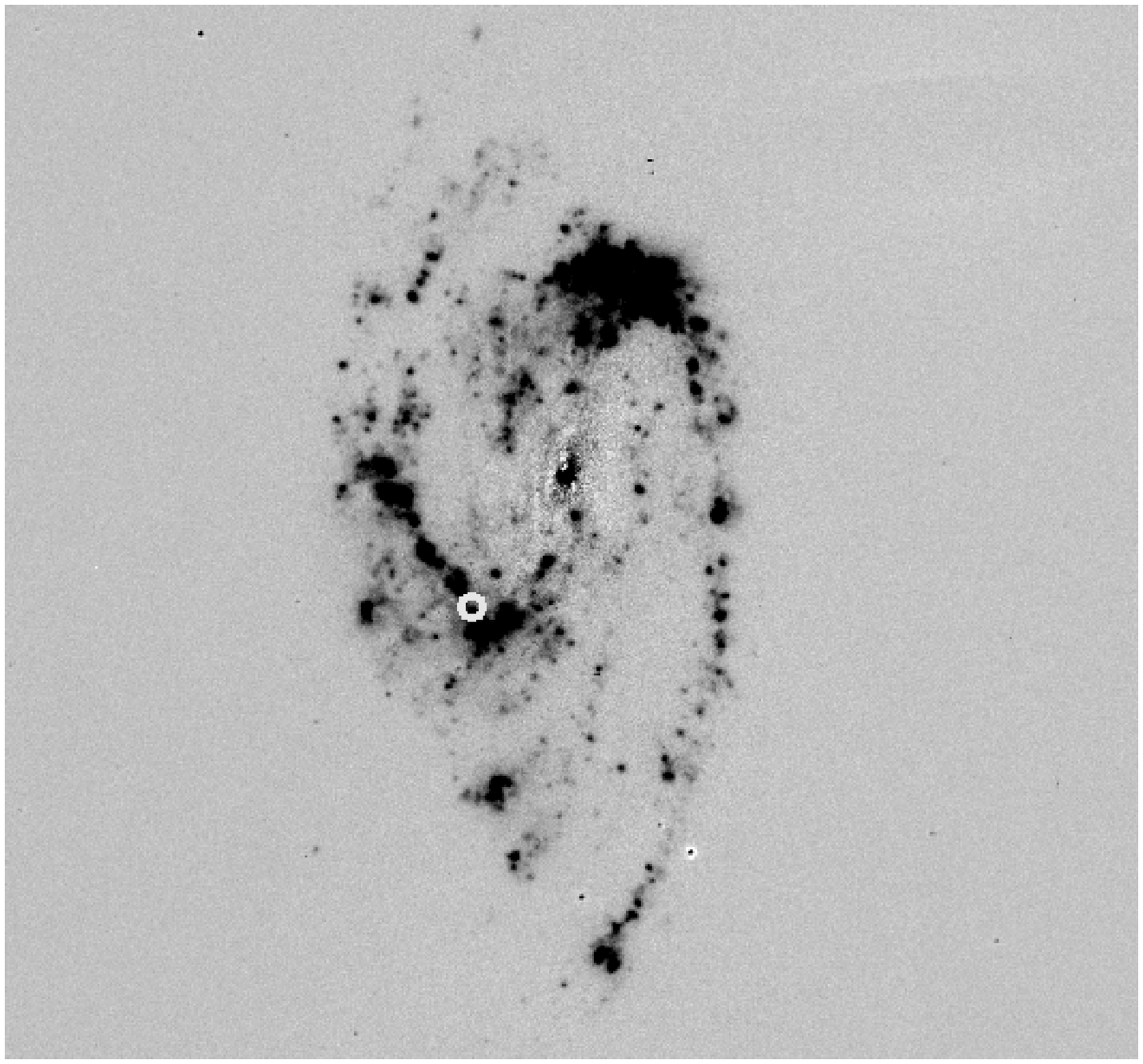}
\caption{Examples of negative continuum-subtracted \ha\ images used
in our analysis. In all images the location of the SN is marked by the white
circle and north is up with east to the left.
The left hand panel shows an ESO 2.2m image of NGC 1433, the host
galaxy of the type IIP SN1985P. Here the `NCR' pixel value is 0.000. In the
centre panel we show an LT image of NGC 1343, the host galaxy of the type Ic
SN2008dv. The analysis of this image gives an `NCR' value of 0.802 for the
SN. Finally in the right hand panel we show an INT image of NGC 3627, the host
galaxy of the type IIL SN2009hd. The `NCR' value for SN2009hd is 0.398.}
\end{figure*}

\section{Data}
\label{data}
The data used for this study have been assembled over a number of years from
range of projects, many of which were not originally focused on SN
environment studies. Therefore our sample is quite heterogenous in terms
of SN and galaxy selection. While this means that the host galaxies analysed
and the relative numbers of SNe contain significant biases, what we should
have is a sample of SNe and their host galaxies which are a random selection
of the, to-date, `observed' local SNe, the only proviso being that we have
favoured SNe with sub-type classification. The data now discussed are an
amalgamation of `new' data with that presented in AJ08.\\ 
All SNe within the sample have host
galaxies with recession velocities less than 10000 \kms\ (the majority 
have recession velocities less than 6000 \kms), with a median 
velocity of 1874 \kms. We exclude all
SNe that occur within highly inclined disk galaxies with axis ratios higher than 4:1, in order
to reduce the occurrence of chance superpositions of SNe onto
foreground/background stellar populations. 
Data that were obtained specifically for this project were chosen where
SNe sub-type information was available in the literature. This introduces the
bias that many of the SNII with sub-type classification fall within galaxies
at lower redshift than the SNIb/c
population. We later discuss the origin of this bias and investigate whether
this has any effect on our results and conclusions. SN types were originally
taken from the Asiago catalogue\footnote{http://graspa.oapd.inaf.it/}, but the
literature was extensively searched for further information, and in some cases
SN types were changed (these instances are listed in table A1). 
We do not include any `02cx-like' objects (see
e.g. \citealt{jha06,fol09}) due to the unclear nature of their origin, 
while we also exclude `Ca-rich' objects (see
e.g. \citealt{fil03,per10,per11}), 
again due a possible distinct progenitor population.  
As we are interested in the stellar population
within the environment of each SN we do not wish to detect any remaining
emission from the SNe themselves. Therefore
we introduce a criterion that host galaxy imaging must have been taken at
least 1 year post SN discovery date for SNIbc and 1.5 years for SNII (which
for the case of SNIIP, the most dominant type, are likely to have longer
lasting light curves). This criterion is difficult to apply to SNIIn. These SNe
sometimes show very long term (sometimes on the order of decades; e.g. \citealt{bau08}) interaction,
which can manifest as \ha\ emission, and thus could easily mimic
environment HII regions where none actually exist. However, given our
results presented in section 4.1 of a \textit{non}-association of SNIIn to SF
as traced by \ha\ emission, any `false' associations, were they removed, would
only increase the significance of our results. Therefore we apply the same
criterion to these SNe as to other SNII.\\
Following AJ08 some of the data initially included in this analysis have now
either been removed or a SN type classification has been changed. Here we
list these changes:\\
\textbf{SN 1996ae:} SN removed from analysis due to axis ratio of host\\
\textbf{SN 2002gd:} SN removed from analysis due to axis ratio of host\\
\textbf{SN 2002bu:} type classification changed to `impostor' following \cite{smi11}\\
\textbf{SN 2004gt:} type classification changed to SNIc following
reclassification in Asiago catalogue\\
\textbf{SN 2006jc:} type classification changed to SNIb following
reclassification in Asiago catalogue\footnote{This SN is actually part of a
peculiar class of Ibc objects which also show signs of interaction with
CSM (e.g. \citealt{fol07,pas08}). However, given that apart from these
signatures 
the SN displays a `normal' Ib spectrum we include it
in our analysis (our results and conclusions would not change if we were to
remove this object).}\\
\textbf{SN 2001co:} SN removed from analysis as classified as `Ca-rich' object
\citep{aaz01}\\
\textbf{SN 2003H:} SN removed from analysis as classified as `Ca-rich' object
\citep{gra03}\\
We note that none of these changes would alter the overall results and
conclusions presented in AJ08.\\
For the subsequent analysis we will use both \ha\ and near-UV emission as 
tracers of SF of different characteristic ages. \ha\ emission observed in SF galaxies is
produced by the recombination of hydrogen ionised by massive stars. This
emission is generally observed as bright HII regions within galaxies that are
thought to be consistent with stellar ages of less than 16 Myr
\citep{gog09}. Near-UV emission (defined here as that detected by the
Galaxy Evolution Explorer; $GALEX$, near-UV pass-band which is centered at
$\lambda$ = 2316\AA) is that produced by hot
massive stars (but including less massive stars than those needed to produce \ha\ emission)
and is thought to trace episodes of SF on timescales between 16 and
100 Myr \citep{gog09}. Hence, \ha\ emission is tracing SF on shorter timescales than
that traced by near-UV emission. Therefore we choose to define SF as traced by
\ha\ as on-going, while that seen at near-UV wavelengths as recent
SF. This nomenclature will be used for the remainder of the paper.

\subsection{New \ha\ imaging}
\label{newha}
\subsubsection{MPG/ESO 2.2m data}
\label{eso2.2}
\ha\ narrow- and $R$-broad band imaging (used for continuum subtraction)
were obtained for 43 CC SNe
host galaxies with the Wide Field Imager (WFI) mounted on the 2.2m MPG/ESO
2.2m telescope (referred to as `ESO' in table A1) at La Silla, Chile, giving 
images with pixel sizes of 0.238 arcseconds per pixel. Data were obtained using the narrow-band
`Halpha/7' and `665/12' filters, and exposure times of $\sim$900s (split into 3x300s to remove cosmic
rays by median stacking) for the narrow band filters and 300s in
$R$-band were generally used. Data
were reduced in a standard way, which will be summarised for all instruments
below.\\
In Fig. 1 we show examples of 3 continuum-subtracted \ha\ images of SN host
galaxies 
used in the current analysis.

\subsubsection{Additional LT data}
\label{LT2}
28 `new' CC SNe were found to have occurred either in old data (i.e. SNe
that were discovered in data presented in AJ08, \citealt{and09}, but after
their publication), or new data specifically obtained for this
project, with RATCam mounted on the Liverpool Telescope (LT) at La Palma, the
Canary Islands. These data were obtained using the narrow \ha\ filter and
broad $r'$-band filter (used for continuum subtraction), and data were binned
2x2 giving pixel scales of 0.276 arcseconds per pixel. Images were obtained
with exposure times of $\sim$900s (split into 3x300s to remove cosmic
rays by median stacking) for the narrow band filter and 300s in the
$r'$-band. These data were automatically processed through the LT pipeline
giving bias-subtracted and flat field-normalisation images that were then
processed as outlined below before analysis.

\subsubsection{Additional JKT data}
\label{jkt}
\ha\ and $R$-band galaxy imaging were recovered from previous projects 
of one of the
authors (PAJ). These were generally projects to investigate the SF properties
of nearby galaxies. The catalogues were searched and 14
`new' CC SNe were found to have occurred within galaxies imaged by the Jacobus
Kapteyn Telescope (JKT), on La Palma, the Canary Islands. 
These images have a pixel scale of 0.33 arcseconds
per pixel. Data were obtained through a range of projects and hence
different filters and exposure times were used for different galaxies. Details
of examples of those data obtained for the \ha GS (\ha\ galaxy survey), and
their reduction, can be found in
\cite{jam04}. Data were generally obtained with exposure times giving similar \ha\ sensitivity as
other data presently analysed. 

\subsubsection{Additional INT data}
\label{int2}
Data were also recovered from other projects that were taken with the Wide
Field Camera (WFC), mounted on the Isaac Newton Telescope, at La Palma, the
Canary Islands. Either new data were found, or additional SNe had occurred within
data presented previously, in 11 cases. Again these data were obtained in a
similar fashion to that above, and were reduced in a similar way. 
The WFC pixel
size is 0.333 arcseconds per pixel.

\subsubsection{\ha\ data reduction}
\label{reduce}
All \ha\ and $R$- (or $r'$-) band data were reduced in a standard manner, which
has been discussed in detail in \cite{jam04} and also in previous papers in
this series (AJ08, \citealt{and09}). The narrow- and broad-band data were
processed through the usual stages of bias-subtraction and flat-field
normalisation before \ha\ images were continuum-subtracted using the
broad-band exposures, using routines in IRAF\footnote{IRAF is distributed 
by the National Optical Astronomy Observatory, which is operated by the 
Association of Universities for Research in Astronomy (AURA) under 
cooperative agreement with the National Science Foundation.} and
\textit{Starlink}. 
The continuum subtraction was achieved by first aligning and scaling the narrow- to
broad-band images, then stars within the field were used to estimate a
continuum scaling flux
factor between the two exposures. After using this factor to normalise the
broad-band images to that of \ha, we used the former to remove the
contribution of the continuum to the narrow-band images, leaving only the flux
produced by \ha\ (and \nii) line emission. In general this process works well,
however in some cases individual images need to be re-processed through an
iteration in order to obtain a satisfactory continuum subtraction.\\
Bright foreground stars often leave residuals after the subtraction
process. These residuals are removed through `patching' of the affected
image regions (pixel values are changed to average values of those just
outside the affected region). When the affected region covers a large portion of
the galaxy, or is extremely close to the SN position 
we remove these cases from our statistics due to the uncertainties
this may cause. Finally all images are processed to leave a mean sky value of zero.\\
In order to accurately derive the position where each SN exploded within its
host galaxy, we require accurate (sub-arcsecond) astrometry. Therefore we
determined our own astrometric solutions for all images within the sample
using comparison of stars with known sky coordinates from second generation
Palomar Sky Survey XDSS images
downloaded from the Canadian Astronomy Data Centre 
website\footnote{http://www4.cadc-ccda.hia-iha.nrc-cnrc.gc.ca/dss/}, 
and used the $Starlink$ routine ASTROM to calibrate images. Finally, all the above images
are binned 3x3 to decrease the effects of uncertainty in SN
coordinates on our analysis. Hence all images are analysed with effective
pixel sizes of $\sim$1 arcsecond. The median recession velocity of the sample
of host galaxies is 1874~\kms. 
Therefore, at this distance we are probing physical
sizes of around 130 pc. At the distance of the closest galaxy 
(NGC 6946) we probe distances of $\sim$40 pc, while for
the most distant galaxy (UGC 10415), the corresponding resolution is 
much coarser, $\sim$650 pc. Hence, in most cases we do not
probe individual HII regions, but larger SF complexes within galaxies. 
Later in
the paper we will show that splitting the sample by host galaxy distance makes
little difference to the results obtained, hence we are confident in our
analysis technique with respect to this issue.

\subsection{GALEX data}
\label{galex}
As we will show below, many CC SNe do not
follow the on-going SF as traced by \ha\ emission. Therefore, we decided to
probe the association of certain SN types with SF of 
longer characteristic lifetimes. To do this we chose to use \textit{GALEX}
near-UV images, and as outlined above we define that the emission detected in
these images is that from recent SF, i.e. that on timescales between 16 and
100 Myr \citep{gog09}. We chose to use near-UV images in
place of \textit{GALEX} far-UV images for two reasons. 1) In terms of the ages
of SF traced by the two, the near-UV emission 
gives a longer time baseline compared to 
that
traced by \ha\ emission, and hence we can hope to see larger differences in
the association of the different SNe to the different emission. 2) There are
fewer detections at far-UV wavelengths and therefore
using the far-UV images would lead to larger uncertainties where no emission
is detected.\\
For all those SNe where we require UV imaging, we use the $GALEX$ search 
form\footnote{http://galex.stsci.edu/GR6/?page=mastform} to download host
galaxy images. $GALEX$ has obtained data for a range of projects from an all
sky survey (AIS), 
to individual time requests of smaller samples (details of the surveys achieved can
be found in \citealt{mor07}). Hence for some galaxies within our
sample there is a range of images to choose from. For each galaxy we use the
deepest images available. These near-UV images display the emission at
wavelengths between 1770 and 2730 \AA\ and have pixel sizes of 1.5 arcsecond
per pixel. Hence, these images probe similar physical sizes to those taken
with ground-based optical detectors as outlined above.\\

\section{Pixel statistics analysis}
\label{pix}
The pixel statistic used throughout this paper has been used and described in detail in previous works. 
Here, we
briefly summarise the formulation and use of this statistic, but we refer the
reader to those previous papers (\citealt{jam06}, AJ08) for more in depth
discussion of the technique and its associated errors. In latter sections we
delve deeper into some un-addressed issues that may be important for the
interpretation of the use of this statistic on our SN and galaxy samples.\\
Our `NCR' (a shortened acronym of the `Normalised Cumulative Rank pixel
function', first presented in \citealt{jam06}) statistic gives, for each
individual SN, a value between 0 and 1 which corresponds to the amount of
emission within the pixel of a SN, in relation to that of the whole host
galaxy. It is formed in the following way, from continuum-subtracted 
\ha\ images (plus \textit{GALEX} near-UV in the present paper). First, the images are
trimmed so that we only include emission of the galaxy and the SN
position. This helps to avoid large fluctuations in the sky background and/or
bright foreground stars in the vicinity of the host, which sometimes 
prove difficult
to remove\footnote{Generally foreground stars are successfully removed
during the continuum subtraction process. As outlined earlier, if the
residuals of these bright stars cover a significant fraction of the host
galaxy area we remove these cases from our results.}. 
The pixels from each image are then ranked in terms of increasing
count; i.e. from the most negative sky value, up to highest count from the
pixel containing the most flux within the
frame. We then form the cumulative distribution of this ranked pixel count,
and finally normalise this to the total flux summed over all pixels. We set all
negative values in this cumulative distribution to 0. Hence, NCR values of 0 
correspond to the SN falling on
zero emission or sky values, while a value of 1 means that the SN falls on the
brightest pixel of the entire image. We proceed with this analysis for all SNe
within the sample (using the same technique for both \ha\ and near-UV images,
where they are included), building distributions for the different CC SN
types. If we assume that the SF pixel count scales by number of stars being
formed (with near-UV emission simply tracing stars down to lower masses), 
then if a SN population accurately follows the stars being formed and mapped  
by that particular SF tracer, we expect that the overall NCR distribution
for that SN type to be flat and to have a mean value of 0.5. We use this as a
starting point for interpreting our results on the different SN types, and
investigate differences in the association of different types with host galaxy
SF.\\ 
The most logical assumption when interpreting these distributions is that
a decreased association to the SF implies longer stellar lifetimes and hence
lower pre-MS masses. Hence we can use this implication to probe differences in
progenitor lifetime and mass of the different CC types. We will discuss the
validity of this interpretation in section 5 and outline how this can be
understood in terms of other progenitor and SF properties. Now we present the
results achieved through this analysis.  

\section{Results}
\label{res}

\begin{figure*}
\includegraphics[width=16cm]{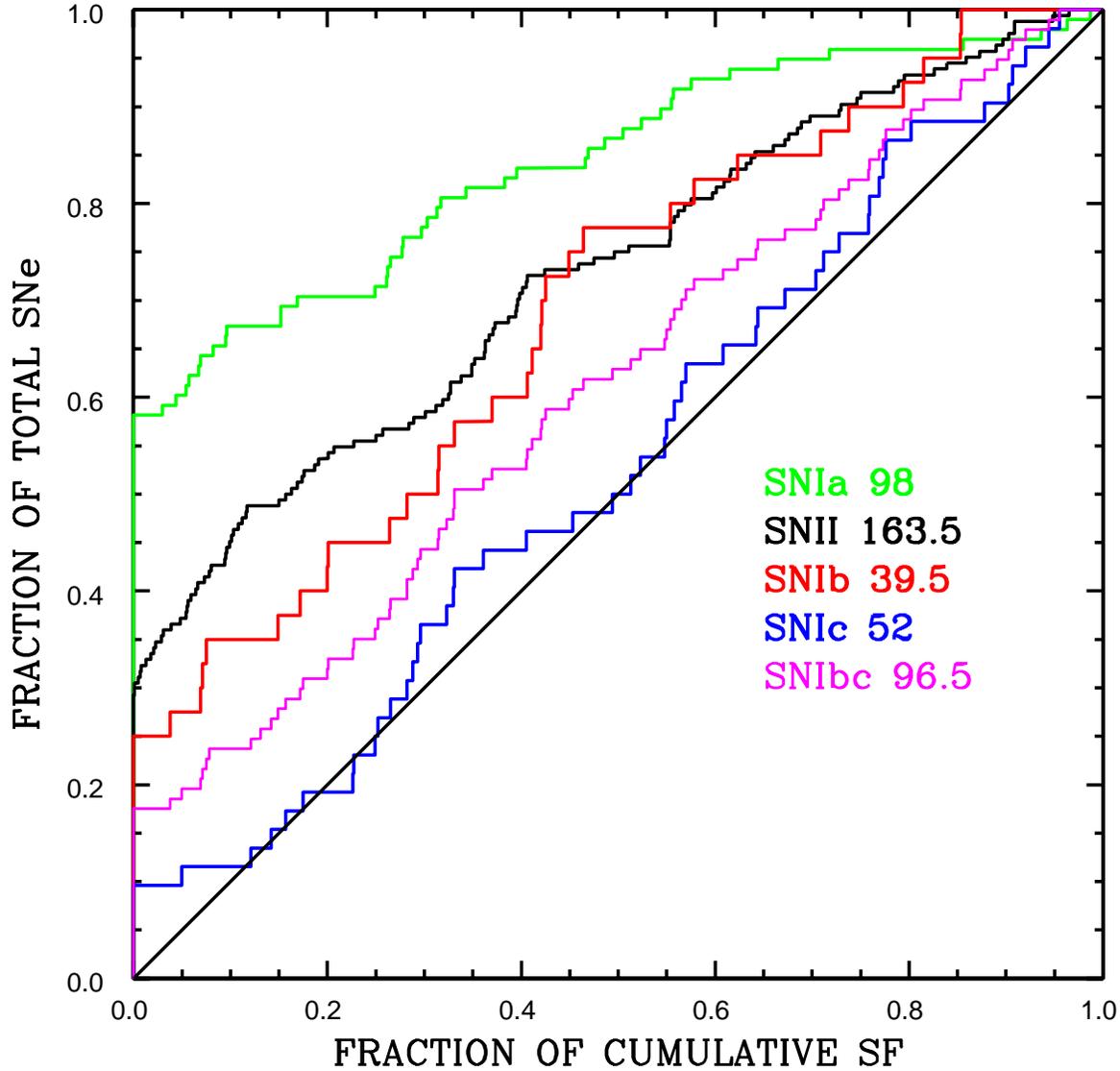}
\caption{Cumulative pixel statistics plot of all the main SN types. SNIa (98
events) are
shown in green, SNII (163.5) in black, SNIb (39.5) in red, the SNIc (52) in blue and the overall
SNIbc (96.5) population in magenta. The black diagonal line illustrates a
hypothetical, infinite in size distribution that accurately follows the
on-going SF. As a distribution moves away to the top left-hand corner from
this diagonal it is displaying a lower degree of association to the
emission. Hence, a clear sequence is displayed, from the SNIa through the
SNII, the SNIb and to the
SNIc in terms of increasing association to the \ha\ line emission.}
\end{figure*}

\begin{table}
\centering
\begin{tabular}[t]{cccc}
\hline
SN type & N & Mean NCR & Std. err.\\
\hline	
\hline
Ia & 98    & 0.157 & 0.026\\
II & 163.5 & 0.254 & 0.023\\
Ib & 39.5  & 0.318 & 0.045\\
Ic & 52    & 0.469 & 0.040\\
\hline
Ibc& 96.5  & 0.390 & 0.031\\ 
\hline
\hline
\end{tabular}
\caption{The NCR pixel statistics for each of the `main' SN types. In the
first column we list the SN type, followed by the number of events within
that distribution in column 2. In column 3 we list the mean NCR value for each
SN type, followed finally by the standard error on that mean. Note, there
are more SNIbc than the sum of the SNIb and SNIc. This is because there are 5
events in our sample where only a `Ib/c' classification is given in the literature.}
\end{table}

The results of the pixel statistics analysis, with respect to
host galaxy on-going SF (as traced by \ha\ emission) are presented in table
A1, along with SN types, galaxy properties and references.\\
First, we present results and distributions for the `main' SN types of SNIa, 
SNII, SNIb and SNIc, while also grouping the SNIbc together
(we present the SNIa distribution here for comparison to the CC SNe, however a full
analysis and discussion of this distribution, splitting the population by
light-curve parameters is being presented elsewhere; Anderson et al., in
preparation). The mean
NCR values for each distribution are presented in table 1 together with
standard errors on the mean and the number of events in each sample. The
distributions of each SN type are displayed in Fig. 2 as cumulative distributions.\\
We use the Kolmogorov-Smirnov (KS) test to probe differences between the
distributions, and also between the distributions and a hypothetical, infinite
in size flat distribution (i.e. one that accurately traces the SF of its host
galaxies). This hypothetical distribution is shown by the diagonal black line
in Fig. 2. The results of these tests are now listed, where a percentage is
given for the likelihood that two populations are drawn from the same
underlying distribution. If this percentage is higher than 10\%\ then we
conclude that there is no statistically-significant difference between the
distributions\footnote{The KS test takes two parameters to calculate this
probability; the `distance' between the two distributions (basically the
largest difference in the y-scale between the distributions as shown in
Fig. 2), and the number of events within each distribution. Hence with small
samples it is hard to probe differences between distributions. Some of the SN
sub-types analysed in this work are dominated by this restriction.}.\\
\textbf{Ia-II:} $\sim$0.1\% \\
\textbf{II-Ib:} $>$10\%  \\
\textbf{Ib-Ic:} $\sim$5\% \\
\textbf{II-Ibc} $\sim$0.5\% \\
\textbf{Probability of being consistent with a flat distribution}\\
\textbf{II-flat:} $<$0.1\% \\
\textbf{Ib-flat:} $<$0.1\% \\
\textbf{Ic-flat:} $>$10\%  \\
\textbf{Ibc-flat:} $\sim$0.5\%\\
We find that SNIa show the lowest degree of association to host galaxy SF of
all SN types, as expected if these SNe arise from WD progenitor stars,
i.e. an evolved stellar population. Following the SNIa we find a sequence of increasing
association to the on-going SF, which implies a sequence of \textit{decreasing}
progenitor lifetime and hence an \textit{increasing} progenitor mass, if we
make the simple assumption that a higher degree of association to SF equates
to shorter `delay-times' (time between epoch of SF and epoch of SN). This
sequence progresses as follows:\\

\begin{center}
\huge{\textbf{SNIa $\Rightarrow$ SNII $\Rightarrow$ SNIb $\Rightarrow$ SNIc}}\\
\normalsize{}
\end{center}

The SNIc appear to arise from the highest mass progenitors of all CC SN
types (indeed higher mass also than any of the other sub-types, given the mean
values presented below). We note that while the SNIc accurately trace the on-going SF
the SNIb do not, while statistically the SNIb show a similar degree
of association to the SF as the overall SNII population.\\
When comparing the overall SNII and SNIbc populations we find that the latter
show a significantly higher degree of association to the \ha\ line
emission. This implies that overall SNIbc arise from more massive progenitors
than the SNII population. We note here that this does not necessarily imply
single star progenitors for SNIbc. Our result solely implies that the SNIbc
arise from shorter lived, higher mass progenitors, whether single or binary
star systems. This issue is discussed in
detail below.  
While all of these results were indicated in our earlier study (AJ08),
the current data set is the first to clearly separate out the SN Ib from the 
SN Ic, with the latter now being seen to be significantly more strongly 
associated with on-going SF, and hence arising from higher mass progenitors.\\

\subsection{CC SN sub-types}

\begin{figure}
\includegraphics[width=8.5cm]{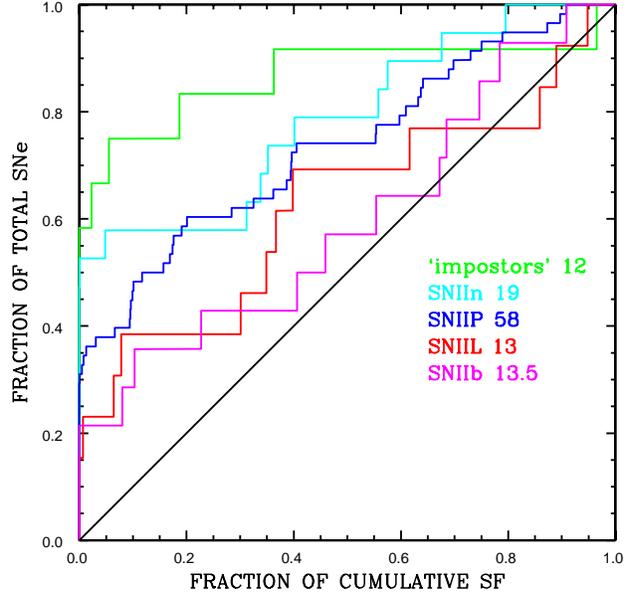}
\caption{Cumulative pixel statistics plot of the CC SNII sub-types. SN
`impostors' (12 events) are shown in green, SNIIn (19) in cyan, SNIIP (58) in
blue, SNIIL (13) in
red, and SNIIb (13.5) in magenta. As in Fig. 2, lines that are further away 
from the
black diagonal line show a lower degree of association to the
on-going SF. Note, the surprising distributions of the SN `impostors' and
SNIIn, both of which lie towards the top left hand side of the figure.}
\end{figure}

\begin{table}
\centering
\begin{tabular}[t]{cccc}
\hline
SN type & N & Mean NCR & Std. err.\\
\hline	
\hline
`impostors' & 12 & 0.133 & 0.086\\
IIn & 19 & 0.213 & 0.065\\
IIP & 58 & 0.264 & 0.039\\
IIL & 13 & 0.375 & 0.102\\
IIb & 13.5 & 0.402 & 0.095\\ 
\hline
\hline
\end{tabular}
\caption{The NCR pixel statistics for each of the SNII sub-types. In the
first column we list the SN type, followed by the number of events within
that distribution in column 2. In column 3 we list the mean NCR value for each
SN type, followed finally by the standard error on that mean.}
\end{table}

We now further separate the CC SN types into various sub-type classifications
that are given in the literature and were discussed earlier in the
paper\footnote{An obvious group to investigate here would be the so-called
`broad-line' class of objects, in particular due to their association to
long-duration GRBs. We searched the literature for evidence of a sample of
these objects within our data but found few compelling cases. Therefore we do
not investigate this group of objects in the present study.}. The mean NCR
pixel values together with their standard errors for the CC sub-types are
presented in table 2, while we show the cumulative distributions of the
different populations in Fig. 3.\\
We perform KS tests between various distributions together with tests
between populations and a hypothetical flat distribution that directly traces
the on-going SF. We now list these probabilities:\\
\textbf{IIb-IIP:} $>$10\% \\
\textbf{IIn-IIP:} $>$10\% \\
\textbf{Probability of being consistent with a flat distribution}\\
\textbf{`impostors'-flat:} $<$0.1\% \\
\textbf{IIn-flat:} $<$0.1\% \\
\textbf{IIP-flat:} $<$0.1\% \\
\textbf{IIL-flat:} $\sim$10\%  \\
\textbf{IIb-flat:} $>$10\% \\
Again we can list these in terms of increasing association to the
\ha\ emission, as is displayed in Fig. 3 (however given the
lower number of events within these distributions, the overall
order of the 
sequence may not be intrinsically correct). We find the following sequence
of increasing association to the line emission, implying a sequence
of increasing progenitor mass:\\

\begin{center}
\textbf{`impostors' $\rightarrow$ IIn $\rightarrow$ IIP $\rightarrow$ IIL $\rightarrow$ IIb}\\
\end{center}

The first observation that becomes apparent looking at this sequence and the
distributions displayed in Fig. 3 is the position of the SN `impostors' and
the SNIIn\footnote{The nature of the transient `1961V' is currently being
debated; whether it was an `impostor' or the final death of a massive star
(see \citealt{smi11,koc11,van12,koc12} for recent discussion). We choose to
keep this event in the `impostor' classification, but we note that moving to
the `IIn' group would make no difference to our results or conclusions. The
NCR value for this object (published in AJ08) is 0.363.}. 
Indeed we find that $>$50\%\ of the SNIIn do not fall on regions of
detectable on-going SF (we will soon evaluate the physical meaning of this
statement). These observations are perhaps surprising given the substantial
literature claims that the progenitors of both of these transient
phenomena are Luminous Blue Variable (LBV) stars. LBVs are massive, blue, hot
stars that go through some extreme mass-loss events \citep{hum94}. These pre-SN
eruptions may provide the CSM needed to explain the signatures of interaction
observed for these transients.
However, these claims are inconsistent with their lower association to SF.\\
Regarding the other sub-types, we find that the SNIIP show a very similar
degree of association to the on-going SF as the overall SNII distribution
presented above. This is to be expected as a) the overall distribution is
dominated by SNIIP and b) it is likely that a large fraction of the SNe simply
classified as `II' (48) in the literature are also SNIIP. The fact
that a large fraction of the SNIIP population does not fall on regions of
on-going SF suggests that large fraction of
these SNe arise from progenitors at the lower end to the CC mass range.\\
The other sub-types included in this study are the SNIIL and SNIIb. Given the
low number statistics involved (13 IIL and 12.5 IIb), definitive results and
conclusions are perhaps premature. However, it may be interesting to note that
both of these types appear to occur within or nearer to bright HII regions
than the SNIIP, implying higher mass progenitors.

\subsection{SNIIP \ha\ vs near-UV}
\label{IIPUV}
The mean NCR value for the 58 SNIIP with respect to \ha\ is
(as above) 0.264 (standard error on the mean of 0.039). This mean value
together with the KS statistic test comparing the population to a flat 
distribution, shows that the SNIIP do not trace the SF as traced by
\ha\ emission. Indeed around 30\%\ of these events fall on pixels containing
no \ha\ emission. 
If we assume that this implies that SNIIP arise from lower mass
massive stars then we expect to see a stronger association to the recent SF
as traced by near-UV
emission. We test this using $GALEX$ near-UV host galaxy imaging and the
resulting 
distribution is
shown in Fig. 4 (for the 50 SNIIP from our sample where near-UV images are
available). We also list the near-UV NCR values in table 3 together with
their \ha\ counterparts and host galaxy information.\\
We find that the SNIIP accurately trace the near-UV emission (although there
are still almost
15\%\ of events that do not fall on regions of SF down to the detection limits
of $GALEX$ in the near-UV). The SNIIP near-UV NCR distribution is formally
consistent (chance probability $>$10\%) with being drawn from a flat
distribution (i.e. accurately tracing the recent SF), while the distributions
with respect to the on-going and recent SF are statistically not drawn from
the same underlying population (KS test probability $<$0.1\%).
As above, this implies that SNIIP arise from the lower mass
end of stars that explode as CC SNe (actual stellar age/mass limits are
discussed below).

\begin{figure}
\includegraphics[width=8.5cm]{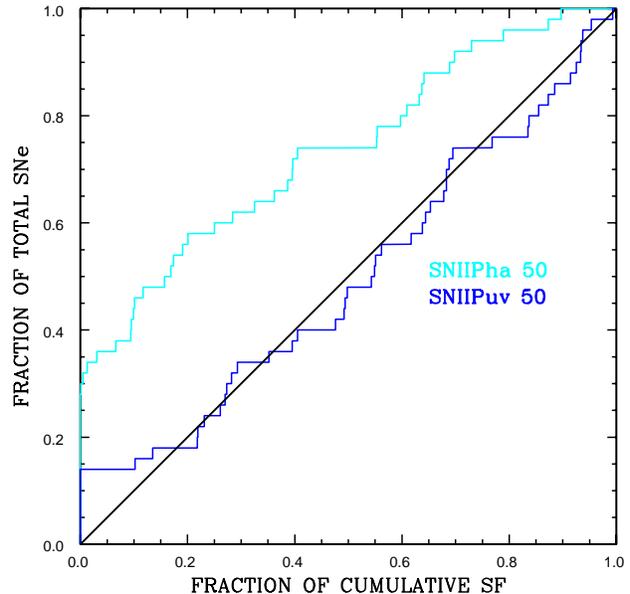}
\caption{Cumulative plot for the pixel statistics of 50 SNIIP with respect to  
\ha\ (cyan) and $GALEX$ near-UV emission (blue). While the SNIIP
do not accurately trace the youngest SF measured by \ha\ line emission they do
follow the near-UV emission which traces SF down to older population ages.}
\end{figure}

\subsection{SNIIn \ha\ vs near-UV}
\label{IInuv}
As discussed above the SNIIn show an even lesser degree of association to the
\ha\ line emission than the SNIIP, with $>$50\%\ of events having an NCR value
of 0. Therefore, as for the SNIIP we re-do the NCR analysis using $GALEX$
near-UV galaxy imaging (using 18 SNIIn with available data). We 
list the near-UV NCR values in table 4 together with
their \ha\ counterparts and host galaxy information.
We find a similar trend to that of the SNIIP; the
SNIIn are formally consistent ($>$10\%) with being randomly drawn from the
distribution of recent SF, although we note a larger number of events (than
the SNIIP) that
fall on regions of zero near-UV emission (22\%\ compared to 14\%\ for the SNIIP). 
While there appears in Fig. 5 an obvious difference between the SNIIn
distributions with respect to the two SF tracers, given the relatively small
number of events in each sample a KS test is not conclusive.
These results imply that
the majority of SNIIn arise from relatively low mass progenitors.\\

\begin{table}
\centering
\begin{tabular}[t]{cccccc}
\hline
SN & Galaxy&V$_\textit{r}$ (\kms )& NCR$_{ha}$ & NCR$_{uv}$\\
\hline
\hline
\textbf{SNIIP} &&&&\\
1936A & NGC 4273 & 2378 & 0.362 & 0.395\\
1937F & NGC 3184 & 592 & 0.000 & 0.000\\
1940B & NGC 4725 & 1206& 0.000 & 0.000\\
1948B & NGC 6946 & 40 & 0.387 & 0.934\\
1965H & NGC 4666 & 1529 & 0.597 & 0.695\\
1965N & NGC 3074 & 5144 & 0.031 & 0.000\\
1965L &	NGC 3631 & 1156 & 0.001 & 0.551\\
1969B &	NGC 3556 & 699 & 0.191 & 0.219\\
1969L &	NGC 1058 & 518 & 0.000 & 0.562\\
1971S &	NGC 493  & 2338& 0.174 & 0.617\\
1972Q &	NGC 4254 & 2407& 0.405 & 0.492\\
1973R &	NGC 3627 & 727 & 0.325 & 0.218\\
1975T & NGC 3756 & 1318 &0.000& 0.102\\
1982F &	NGC 4490 & 565 & 0.095& 0.549\\
1985G &	NGC 4451 & 864 & 0.641& 0.933\\
1985P &	NGC 1433 & 1075 & 0.000 & 0.135\\
1986I &	NGC 4254 & 2407& 0.000& 0.914\\
1988H &	NGC 5878 & 1991& 0.000& 0.000\\ 
1989C &	UGC 5249 & 1874& 0.689& 0.993\\
1990E &	NGC 1035 & 1241& 0.000& 0.644\\
1990H &	NGC 3294 & 1586& 0.000& 0.270\\
1991G &	NGC 4088 & 757 & 0.066 & 0.231\\
1997D &	NGC 1536  & 1217 & 0.000 & 0.000\\
1999bg &IC 758 &  1275 & 0.632& 0.937\\
1999br &NGC 4900 & 960& 0.099& 0.273\\
1999gi &NGC 3184 & 592& 0.637 & 0.953\\
1999gn &NGC 4303 & 1566& 0.897 & 0.873\\
2001R &	NGC 5172 & 4030& 0.000 & 0.000\\
2001X &	NGC 5921 & 1480 & 0.698 & 0.543\\
2001du &NGC 1365 & 1636& 0.101 & 0.837\\
2001fv &NGC 3512 & 1376& 0.169 & 0.282\\
2002ed &NGC 5468 & 2842& 0.395& 0.683\\
2002hh &NGC 6946 & 40 & 0.000& 0.653\\
2003Z &	NGC 2742 & 1289& 0.013& 0.688\\
2003ao &NGC 2993 & 2430 & 0.157& 0.498\\
2004cm &NGC 5486 & 1390 & 0.201 & 0.638\\
2004dg &NGC 5806 & 1359 & 0.554& 0.855\\
2004ds &NGC 808 & 4964& 0.250& 0.678\\
2004ez &NGC 3430 & 1586& 0.094&0.293\\
2005ad &NGC 941 & 1608 & 0.000& 0.261\\
2005ay &NGC 3938 & 809 & 0.873 & 0.683\\
2005cs &NGC 5194 & 463 & 0.396 & 0.768\\
2005dl &NGC 2276 & 2416 & 0.730& 0.835\\
2006my &NGC 4651 & 788 & 0.553 & 0.476\\
2006ov &NGC 4303 & 1566 & 0.284& 0.885\\
2007aa &NGC 4030 & 1465 & 0.117& 0.352\\
2007od &UGC 12846 & 1734 & 0.000& 0.000\\
2008M &	ESO 121-g26 & 2267& 0.789 & 0.925\\
2008W &	MCG -03-22-07 & 5757& 0.005& 0.405\\
2008X &	NGC 4141 & 1897& 0.609 & 0.494\\
\hline				       
\hline
\end{tabular}
\caption{List of data for all SNIIP where $GALEX$ near-UV host galaxy images
were available in the archive. In column 1 we list the SN name, followed by
its host galaxy and recession velocity in columns 2 and 3. We then list
the \ha\ NCR values followed by their corresponding near-UV values.}
\end{table}

\begin{figure}
\includegraphics[width=8.5cm]{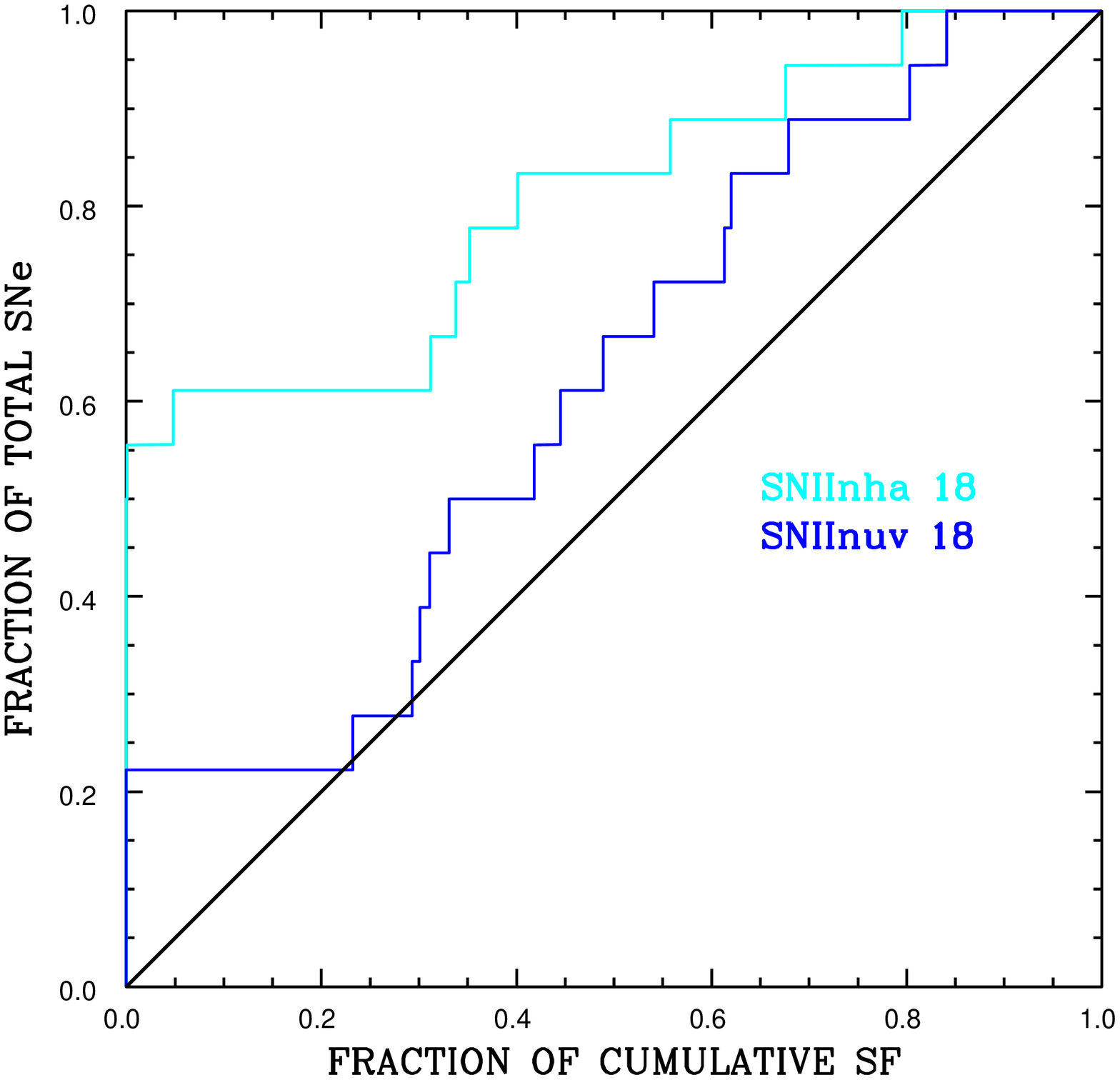}
\caption{Cumulative plot for the pixel statistics of 18 SNIIn with respect to  
\ha\ (cyan) and $GALEX$ near-UV emission (blue). While the SNIIn
do not accurately trace the youngest SF measured by \ha\ line emission they do
follow the near-UV emission which traces rather older populations.}
\end{figure}

\begin{table}
\centering
\begin{tabular}[t]{cccccc}
\hline
SN & Galaxy&V$_\textit{r}$ (\kms )& NCR$_{ha}$ & NCR$_{uv}$\\
\hline
\hline
\textbf{SNIIn} &&&&\\
1987B & NGC 5850 & 2556 & 0.000 & 0.000\\
1987F &	NGC 4615 & 4716 & 0.352 & 0.541\\
1993N &	UGC 5695 & 2940 & 0.000 & 0.000\\
1994Y &	NGC 5371 & 2558 & 0.000 & 0.331\\
1994W &	NGC 4041 & 1234 & 0.795 & 0.679\\
1994ak&	NGC 2782 & 2543 & 0.000 & 0.311\\
1995N &	MCG -02-38-17 & 1856&0.001 & 0.000\\ 
1996bu&	NGC 3631 & 1156 & 0.000& 0.000\\ 
1997eg&	NGC 5012 & 2619 & 0.338& 0.418\\
1999el&	NGC 6951 & 1424 & 0.048 & 0.232\\
1999gb&	NGC 2532 & 5260 & 0.676 & 0.489\\
2000P &	NGC 4965 & 2261 & 0.393 & 0.620\\
2000cl&	NGC 3318 & 2775 & 0.312 & 0.613\\
2002A &	UGC 3804 & 2887 & 0.401 & 0.803\\
2002fj&	NGC 2642 & 4345 & 0.558 & 0.841\\
2003dv&	UGC 9638 & 2271 & 0.000 & 0.301\\
2003lo&	NGC 1376 & 4153 & 0.000 & 0.293\\ 
2006am&	NGC 5630 & 2655 & 0.000 & 0.445\\
\hline				       
\hline
\end{tabular}
\caption{List of data for all SNIIn where $GALEX$ near-UV host galaxy images
were available in the archive. In column 1 we list the SN name, followed by
its host galaxy and recession velocity in columns 2 and 3. We then list
the \ha\ NCR values followed by their corresponding near-UV values}
\end{table}

\subsection{SN `impostors' \ha\ vs near-UV}
\label{impnuv}
Finally we repeat this analysis of \ha\ against near-UV pixel statistics for
the SN `impostors' (11 events). Again we find similar trends that the population shows a
higher degree of association to the recent than the on-going SF. We 
list the near-UV NCR values in table 5 together with
their \ha\ counterparts and host galaxy information. Given the 
low number statistics it is hard to fully trust the results. However, the two
distributions as displayed in Fig. 6 suggest that while the `impostors'
show a higher degree of association to the near-UV compared to the
\ha\ emission, they are not accurately tracing the recent SF. Indeed, 
using a KS test we find only $\sim$3\%\ chance probability that
these events are drawn from the underlying distribution of near-UV emission. This
either implies that these events arise from much lower mass progenitors than
all other transients studied here, that there is
some strong selection effect against finding these events within bright HII
regions, or that the stellar birth processes of these objects differ from those which
form `normal' massive stars.\\
We have not chosen to pursue investigations of the correlation of all other types 
with near-UV emission. We chose the SNIIP, IIn and `impostors' for study because these
were the obvious events of interest, given
their non-association to the \ha\ emission. 

\begin{figure}
\includegraphics[width=8.5cm]{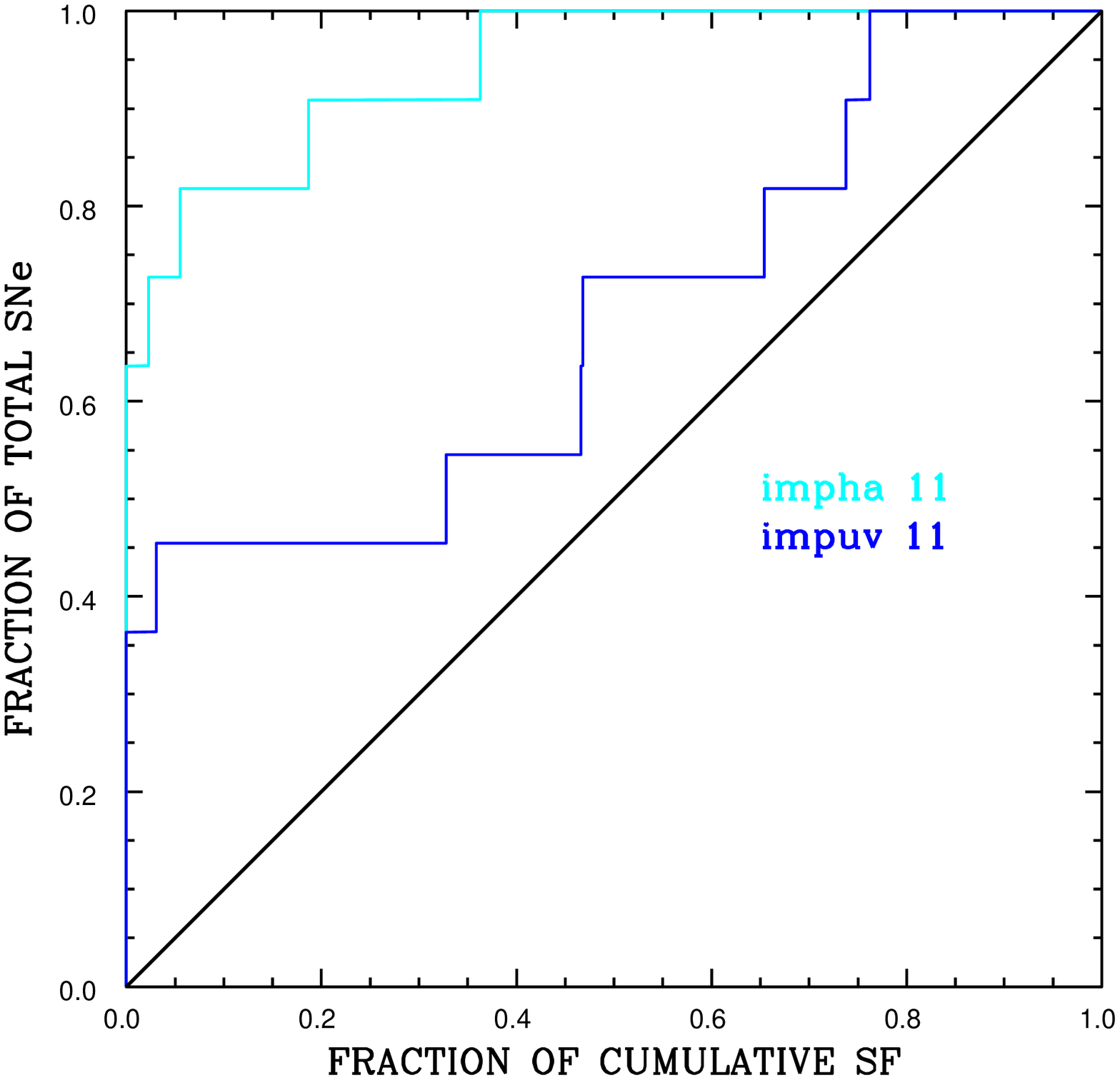}
\caption{Cumulative plot for the pixel statistics of 11 SN `impostors' with respect to  
\ha\ (cyan) and $GALEX$ near-UV emission (blue). The `impostors'
do not accurately follow the youngest SF measured by \ha\ emission, and while
they display a higher association to the near-UV emission they do not appear
to accurately trace it.}
\end{figure}

\begin{table}
\centering
\begin{tabular}[t]{cccccc}
\hline
SN & Galaxy&V$_\textit{r}$ (\kms )& NCR$_{ha}$ & NCR$_{uv}$\\
\hline
\hline
\textbf{SN `impostors'} &&&&\\
1954J&	NGC 2403  & 131 & 0.187 & 0.738\\
1961V &	NGC 1058 & 518 & 0.363 & 0.000\\
1997bs&	NGC 3627 & 727 & 0.023 & 0.328\\
1999bw&	NGC 3198 & 663 & 0.000 & 0.466\\
2001ac&	NGC 3504 & 1534& 0.000 & 0.000\\
2002bu&	NGC 4242 & 506 & 0.000 & 0.000\\
2002kg&	NGC 2403  & 131 & 0.055 & 0.654\\
2003gm&	NGC 5334 & 1386& 0.000 & 0.468\\
2006fp&	UGC 12182 &1490& 0.000 & 0.000\\
2008S &	NGC 6946 & 40 & 0.031 & 0.000\\
2010dn&	NGC 3184 & 463 & 0.000 & 0.762\\
\hline				       
\hline
\end{tabular}
\caption{List of data for all SN `impostors' where $GALEX$ near-UV host galaxy images
were available in the archive. In column 1 we list the SN name, followed by
its host galaxy and recession velocity in columns 2 and 3. We then list
the \ha\ NCR values followed by their corresponding near-UV values}
\end{table}

\subsection{Progenitor age and mass constraints}
\label{constrain}
The most robust results presented here are the relative
differences between the distributions of the different SN types with respect
to host galaxy SF. 
However, given estimates of the SF ages traced by the two wave-bands
used in our analysis, we can go further and make some
\textit{quantitative} constraints. \\
Earlier, we defined the \ha\ to be tracing the on-going SF, on timescales of
less 16 Myr, while defining the near-UV to be tracing recent SF on timescales
of 16-100 Myr (both timescales are taken from \citealt{gog09}). Hence, we can
use these timescales to quantitatively constrain the ages (and hence masses)
of different progenitors. We do this by simply asking if a SN distribution
accurately traces (i.e. a KS test between a SN population and a `flat'
distribution gives a probability of $>$10\%)
either
the on-going or recent SF and then apply the above timescales to the
results. We then use table 1 from \cite{gog09}, taken from the models of
\cite{mar08} to obtain progenitor mass constraints. As above, the constraints
we present below are based on the assumption that a decreasing association to
SF equates to longer lived, less massive progenitors.\\
Before proceeding we need to make an important caveat to the results presented in this subsection. 
While it is generally accepted that near-UV emission traces SF on timescales 
longer than that of \ha, the definitive timescales we present above are less secure. 
While it may be that stars with ages $\sim$16 Myr produce a small amount of ionizing
flux, contributing to that which produce HII regions, the flux will be dominated 
by the most massive stars. Hence, it has been argued (P. Crowther, private 
communication 2012) that only the most massive stars will be found to reside within 
large HII regions. Therefore it could be that only stars with much shorter timescales
than 16 Myr accurately trace the spatial distribution of \ha\ emission within galaxies.
However, these much higher mass (shorter lifetime) values have not been convincingly
shown observationally. Hence, in this section we continue with the age (and therefore
mass) ranges that we take from \cite{gog09} (which were for one
environment within one SF galaxy). Finally, we stress that our
main results and conclusions are not dependent on the strength of these age limits.\\
The positions of SNIIP within host galaxies show that they accurately trace
the recent and not the on-going SF. Hence we can put an upper limit 
for the \textit{majority} of SNIIP to arise from progenitors with ages $>$ 16 Myr
and  masses $<$ 12\msun\ (given the age to mass conversions in table 1 from
\citealt{gog09}). 
We note that the upper mass constraint 
does not exclude
progenitor masses above 12\msun, just that the majority will be produced from
stars below this mass. Given the shape of the IMF (e.g. \citealt{sal55}), even
if the \textit{possible} range of progenitors extends out to above
20\msun\ we still expect the majority to be from the low mass end. This is
consistent with the direct detections thus far for SNIIP (see
\citealt{sma09b} for a review).\\
Moving next to the SNIc, these events accurately trace the on-going SF and
hence the shortest lived, most massive stars. Using the timescale above we
constrain these progenitors to be above 12\msun, hence more massive than the
SNIIP. 
Given that the masses of these SNe
\textit{if} they arise from single stars are probably more than 25\msun\ (from
observational upper limits of red supergiants; \citealt{lev07}, and
predictions from single star models; e.g. \citealt{heg03,eld04,geo09}), this
constraint does not allow us to differentiate between single and binary star scenarios.\\
In Fig. 2 we see that the SNIb fall in between the SNIc and SNII in terms of
their association to the on-going SF. Indeed they are inconsistent with
being drawn from the \ha\ emission distribution. As with the SNIIP, \textit{if}
we take the limits for the production of HII regions detected through \ha\ as 16 Myr 
then this would constrain the \textit{majority} of SNIb progenitors
to be stars less massive than $\sim$12\msun. Keeping the insecure nature of this mass limit in mind,
this would then constrain SNIb to arise from binaries (e.g. \citealt{pod92}), as single stars
less massive than $\sim$25\msun\ are not thought to be able to lose their hydrogen 
envelopes prior to SN (e.g. \citealt{heg03,eld04,geo09}).\\
Given the small statistics for the SNII sub-types, quantitative constraints are
more difficult. Therefore we only apply this argument to the SNIIn. The SNIIn  
do not (KS probability $<$0.1\%) trace the on-going SF, and hence as
for the SNIIP, this argues that the majority of the events within our sample
had progenitor stars with ages of more than 16 Myr and therefore masses of
less than 12\msun.

\subsection{SNe falling on `zero' star formation}
\label{label}
\begin{figure}
\includegraphics[width=8.5cm]{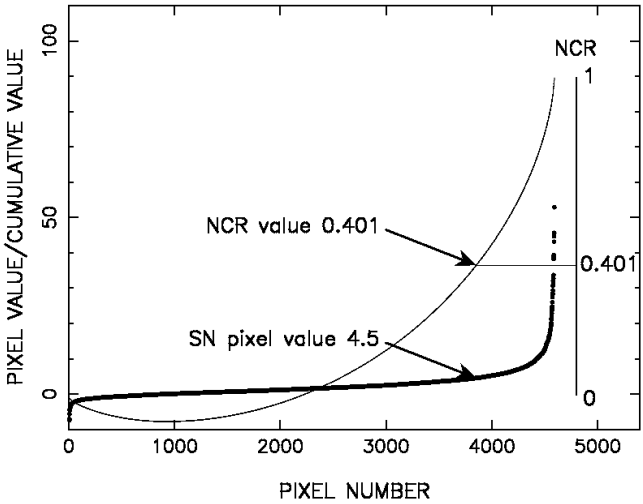}
\caption{Example of the formation of the NCR pixel statistic (figure
reproduced from \citealt{jam06}). The thicker black curve shows the
distribution of increasing pixel count, while the thinner curve shows the
corresponding cumulative distribution. On the right a scale is shown
to represent the NCR values of these distributions. We show this here to draw attention to how
this statistic is formed with respect to sky or zero detected emission
pixels. 
}
\end{figure}

The above analysis has shown that many CC SNe do not fall on regions of
on-going and/or recent SF. Here we investigate this
further and determine whether this is due to our detection
limits or whether there is a significant fraction of events that do indeed
explode away from HII regions. This can be achieved by
evaluating SN pixel values and determining where these fall in the overall
NCR pixel distribution. In Fig. 7 (taken from \citealt{jam06}) we show an
example of how the NCR statistic is formed in relation to the rank of the
pixels of the host galaxies and their cumulative distributions.\\
The NCR statistic is formed by ranking all pixels of the host galaxy,
including those from the surrounding sky, into a sequence of increasing pixel
count. From this count the cumulative distribution is formed. The NCR
statistic is considered to have a non-zero value when this distribution
becomes positive.  
In Fig. 7, this occurs where the thin line crosses the cumulative value
of zero, where the thick line of individual pixel values has a small but 
non-zero positive value.
Therefore there are many positive pixel counts that have an NCR value of
zero. Hence, in forming the NCR statistic we are effectively putting
a limit per pixel for detection of emission. This means that there will be
pixels which correspond to an NCR value of zero \textit{but} contain emission,
just emission that falls below our detection limit.
Now, we can evaluate the cases where a SN has an NCR value of zero
and see where this lies within the pixel distribution that is
considered as zero. If indeed these SNe
are arising from regions consistent with zero intrinsic SF then
the distribution of these NCR=0 pixel counts should be evenly distributed
either side of zero (the mean sky flux in this statistic).\\
We do this analysis for the 15 SNIIP that fall on zero \ha\ NCR
values. We determine the pixel count on which the
SN falls and normalise this so that for each SN we have a value
between -1 and 1, where a value of -1 means that the SN falls on the most
negative count of the image, and a value of 1
means that the SN falls on the most positive count before the NCR
distribution becomes non-zero. We plot
the resulting distribution in Fig. 8. We find that the distribution is
indeed biased towards
positive pixel values. If the SNe were consistent with being drawn from sky or
true zero emission pixels we would expect the same number to have negative and
positive pixel values within the NCR=0 distribution. Given that there are 3 SNe
with negative values  
and 12 with positive values we find an excess of 9 SNe which fall
on zero NCR values \textit{but} fall on emission which is below our detection
limits.\\
There is no way of knowing which of these 12 positive values are the 9 which
fall on intrinsic emission. To estimate the detection limits per
pixel of our imaging (and hence our NCR statistic), we estimate SF rates
(SFRs) per SN containing pixel. For 5 of the galaxies SFRs have been published in
\cite{jam04}. 
For these we take the count for the SN containing pixel and
divide this by the overall galaxy \ha\ count. This then gives us a scaling
factor to apply to the SFR of the host which we use to
calculate a SFR for the SN containing pixel. 
We use these 5 `calibrated' galaxies as representative of the
overall sample of 12 galaxies (this is reasonable given that the SN to total 
pixel counts ratio is very similar between the `calibrated' and `uncalibrated' samples). 
The mean SFR in SN containing pixels is 
2.1$\times$10$^{-5}$
solar masses per
year. We assume this to be our NCR median detection limit for our
\ha\ imaging. 
Hence, when we are talking about a certain
fraction of a SN population that fall on `zero NCR' values, we are talking
about the fraction of events that fall on pixels consistent with (on average) SFRs
of around, or below 
2.1$\times$10$^{-5}$
\msun\ per year. For reference the SFR for the Orion nebula is $\sim$7.9$\times$10$^{-5}$
(taking the \ha\ luminosity from \citealt{ken84}, and converting to a SFR
using eq. 2 from \citealt{ken98}). Therefore, within the median of our SNIIP host galaxies, 
we would expect to detect SF regions of the size/luminosity of Orion, but not those
with smaller SFRs. 
\begin{figure}
\includegraphics[width=8.5cm]{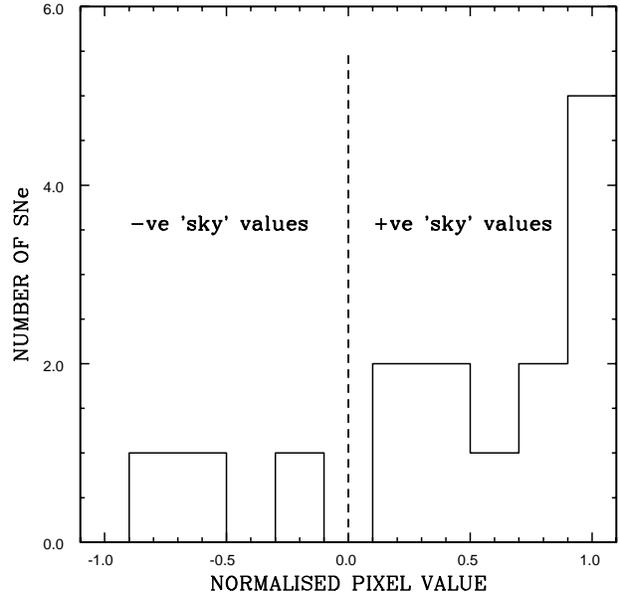}
\caption{Histogram showing the distribution of the normalised pixel values for
the 15 SNIIP which have \ha\ NCR value equal to zero. If these SNe really were falling on regions of zero
intrinsic \ha\ emission then we would expect this distribution to be evenly distributed
between -1 and 1. As the displayed distribution is clearly biased towards
positive pixel values this shows that some of these SNe fall on
emission which is not detected in our imaging.}
\end{figure}
A small number of SNe (IIP, IIn and `impostors') also fall on zero near-UV NCR
values. 
Hence, we
ask the same question as above, whether these are really SNe
that are falling on zero SF or whether this is simply a detection sensitivity issue.\\
On inspection of the images used it is clear that we cannot form the
same distribution as displayed in Fig. 8. This is because the vast majority
of $GALEX$ `sky' pixel counts are simply zero (meaning that when one subtracts the 
mean sky value an image is left with many negative count pixels of the same value). 
Duplicated values are unhelpful for proceeding with the analysis presented above for \ha. 
Therefore, here we simply
calculate whether the pixel where each SN falls is positive or
negative.\\
For the combined sample of SNIIP, IIn and `impostors' there are 16 SNe which fall 
on zero near-UV NCR pixels. We find that 10 have positive sky values and 6 have 
negative counts.
Taking the number of `negative' SNe as 6 means that there will also
be 6 `positive' SNe that are falling on regions consistent with zero
emission. 
Therefore there are only 4 SNe which do
fall on regions of emission, but that which is below our detection
limits. We conclude that there is a small \textit{but not
insignificant} number of CC SNe which fall on regions of zero
\textit{intrinsic} SF as traced by \ha\ and $GALEX$ near-UV
emission.\\
One may ask the question of whether the differences between the degree of
association of the different SN types to host SF are merely due to the overall
number of events which fall on NCR=0 pixels. Indeed, when we look at Fig. 2
we see both a sequence of populations moving away from the flat
distribution, while also seeing a sequence of NCR=0 fractions when
looking only at the y-axis (the position where each distribution starts from the
left hand side of this plot gives the fraction of events within the
distribution that fall on NCR=0 pixels). To investigate whether additionally
there are differences in the
shape of each distribution, we re-plot the populations removing these NCR=0
objects. The resulting cumulative distributions are shown in Fig. 9. We see
that there still appears to be a sequence of increasing association to the
on-going SF within these samples. The KS test of the modified
SNII and SNIc distributions indicates that these are still significantly 
different, at the 2.5\% level, while both the SNII and SNIb
populations are still not consistent with being drawn from a flat 
distribution (probabilities of
$<$0.1\%\ and $\sim$5\%\ respectively). We therefore observe that SNII and SNIb
show both a higher fraction of events falling on zero on-going SF 
\textit{and} less frequently explode within bright HII regions than SNIc.

\begin{figure}
\includegraphics[width=8.5cm]{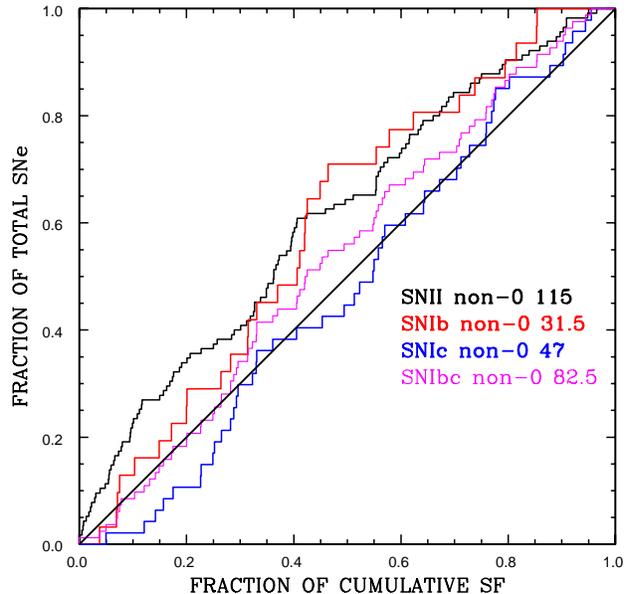}
\caption{Cumulative pixel statistics with respect to \ha\ host galaxy emission
for the main CC SN types, with those NCR
values equal to zero removed. SNII (115 events) are shown in black, SNIb (31.5)
in red, SNIc (47) in blue, and the overall SNIbc sample (82.5) in magenta. We
see that even when these zero values are removed differences remain between
the distributions.}
\end{figure}

\subsection{Selection effects and possible sample biases}
\label{bias}
In this section we test for several possible biases that may be affecting our
results. The main result from this paper is, that
SNIbc show a higher degree of association to host galaxy on-going SF than
SNII. 
The SNII sample is dominated by
SNIIP. One possible source of concern is that the peak brightness 
luminosity function of SNIIP
is seen to extend to lower values than SNIbc. While this difference is
observed, 
the mean luminosities for the IIP and Ibc
samples
presented by \cite{li11} are very similar, and the only 
differences in the distributions concern a faint tail of $\sim$ 30 per cent of SNII. 
Nevertheless SNIIP \textit{may} be harder to detect against bright HII regions, and
be affected by a selection effect.\\ 
If this is strongly affecting our sample then we would expect
to see two trends. Firstly, we would expect the SNII to
SNIbc ratio to decrease with increasing distance. This is because as one goes
to larger distances it becomes harder to detect objects against the background
galaxy emission. Given the (\textit{possible}) luminosity differences between the two samples we
may then expect this effect to be worse for the SNIIP than the
SNIbc. Secondly, we would expect that for both SNIIP and SNIbc as one
goes to larger distances it would become harder to detect both
sets of events against bright HII regions. This would then manifest itself as 
decreasing NCR pixel values for each sample. 
In the
analysis and discussion below we test these hypotheses.\\ 
\begin{table}
\centering
\begin{tabular}[t]{cccc}
\hline
SN type & N & Mean Vr & Median Vr\\
\hline	
\hline
II & 163.5 & 2183 & 1566\\
Ibc & 96.5 & 2833 & 2513\\
\hline
IIP & 58 & 1697 & 1484\\
IIL & 13 & 1285 & 1238\\
IIb & 13.5&2526 & 2426\\
IIn & 19 & 2779 & 2619\\
imp & 13& 852 & 628\\
Ib & 39.5&2809 & 2798\\
Ic & 52& 2919& 2443\\
\hline
II (no sub-type) &47& 3080 & 2407\\
\hline
\hline
\end{tabular}
\caption{Mean and median recession velocities for the host galaxies of the
different SN types. 
In the first colum the SN group is listed, followed
by the number of events within that group in column 2. In column 3 and 4 we
list the mean and median host galaxy recession velocities}
\end{table}
In table 6 we list the mean and median recession velocities (which equate
to distances) of the host galaxies of each sample and sub-sample of SNe
analysed within this work. This initial analysis shows that the SNII
sample is nearer in distance than that of the SNIbc. This trend is
also shown in Fig. 10. However, there is an obvious target selection bias which
is affecting these distributions. Namely, as outlined in section 2, our
target SNe host galaxies were chosen to give SNe with sub-type
classifications. To classify a SN as a Ibc one needs only a
spectrum. However, to classify a SN as a bona-fide IIP or IIL one needs a
light-curve of reasonable quality. The community is more likely to take the
photometry needed to make this distinction for nearby events. Indeed, apart
from the `impostors' the SNIIP and SNIIL have the closest host galaxies, while
the events only classified as `II' have considerably larger distances. Adding this to
the fact that SNIbc are simply rarer and therefore one has to go to further
distances to compile a significant sample and this explains the trend seen in
Fig. 10.\\  
\begin{figure}
\includegraphics[width=8.5cm]{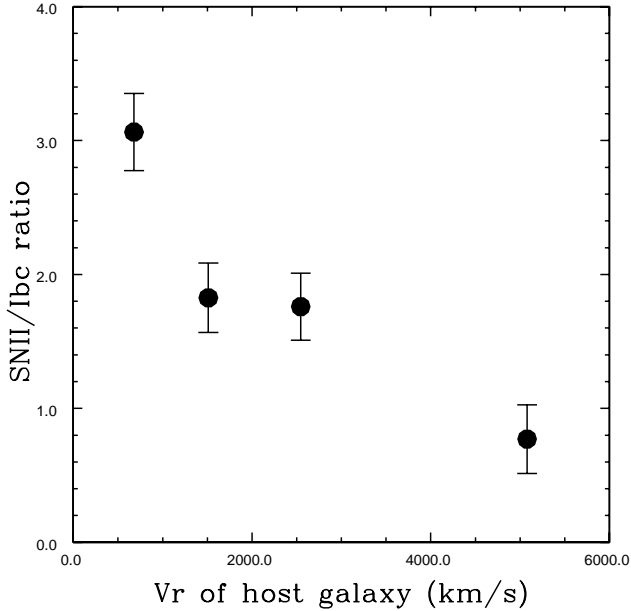}
\caption{The SNII to SNIbc ratio as a function of host galaxy recession
velocity. The data is simply split into quartiles of recession velocity and
the ratio of events is calculated for each bin. Error bars are calculated using
Poisson statistics.}
\end{figure}
Next we split both the SNIIP and SNIbc samples into 4
equal bins of recession velocity. For each of these bins we calculate the mean
NCR value. We then plot these NCR distributions in Fig. 11. While we see the
obvious offset in recession velocities of the SNIIP and SNIbc described
above, both distributions appear to be very flat with host galaxy
velocity. Hence, we conclude that there is \textit{no selection effect which
preferentially detects SNIbc within bright HII regions with respect to
SNII}.\\
Armed with this last conclusion we now discuss all the above results in more
detail, confident that we are seeing true \textit{intrinsic} differences in the
association of different SNe with host galaxy SF.

\begin{figure}
\includegraphics[width=8.5cm]{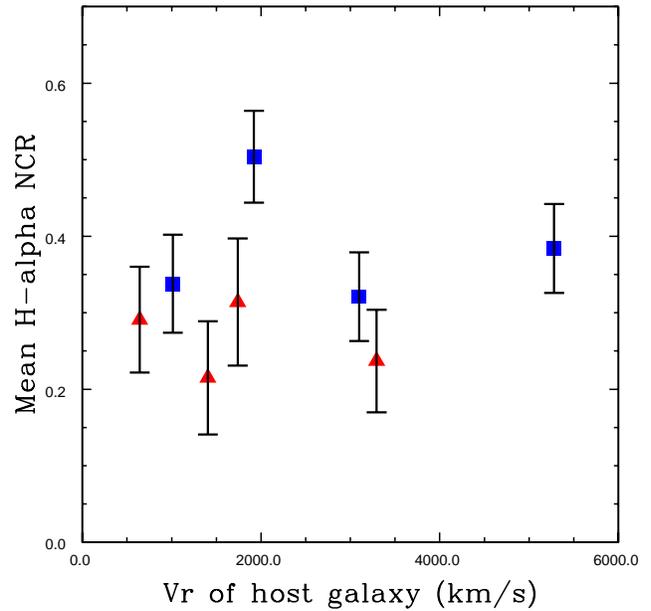}
\caption{The distribution of mean \ha\ NCR values for the SNIIP and SNIbc
samples, split into quartiles of recession velocity. The SNIIP are shown in
red triangles while the SNIbc are shown in blue squares. Error bars are the
standard error on the mean of each sub-sample.}
\end{figure}

\section{Discussion}
\label{diss}
The major assumption employed in this work is that an increasing association
to SF equates to shorter
pre-transient lifetimes. This would appear to be the most logical way to interpret
these results. Here we delve deeper into the physical
causes of these associations and how one can interpret these in terms of
various parameters at play within SN host environments. To motivate this
discussion we start by outlining different reasons a SN will occur at
different distances from HII regions within hosts.\\
\textbf{1)} Only very massive stars (i.e. larger than 15-20\msun, \citealt{ken98,gog09}) produce
sufficient quantities of ionizing flux to produce a bright HII region
visible as \ha\ line emission. When these stars explode as SNe the HII region
from an episode of SF ceases to exist, and hence when lower mass stars from
the same SF episode go SN they will do so in an environment devoid of
emission. Therefore, the higher mass progenitors will better trace the on-going SF
than their lower mass counterparts.\\
\textbf{2)} Longer lived, lower mass stars have more time to drift away from
their places of birth, i.e. HII regions. Therefore, these lower mass
progenitors will explode in regions of lower SF density and overall will trace
the SF to a lesser degree. This scenario is also important for continuous
SF within the same environment. We can envisage continuous SF where
SNe from an initial SF episode trigger further formation
events. In this scenario the most massive stars have little time
pre-explosion to move away from their host HII regions, while lower mass
events have an increasingly long duration of time to drift away.\\
\textbf{3)} Hydrogen gas which is ionized to produce bright HII regions
(through recombination) is blown away by winds from the most massive stars, and their
subsequent explosions. 
Hence, even if lower mass stars still have the necessary ionizing flux
to produce HII regions, there is simply not enough gas within the
environment. 
Therefore, while the most massive stars are observed to explode
within a dense SF region, the lower mass stars explode into a sparse region
devoid of \ha\ emission, again producing differences in the degrees of
association of different mass stars with HII regions.\\
\textbf{4)} Stars that are found away from bright HII regions can be `runaway'
events with high velocities, and have as such moved considerable distances between
epoch of SF and epoch of SN. Lower mass stars are more likely to be influenced by
this effect through both the binary-binary \citep{pov67} and supernova
\citep{bla61} proposed formation mechanisms, due to preferential ejection of
the lowest mass star in the former, and the lower mass star being `kicked' by
the explosion of the higher mass companion in the latter (indeed this scenario
has been addressed in detail for SNe progenitors; \citealt{eld11}). 
Hence, as above, SNe produced by lower mass progenitors
would be found to occur further away from SF regions than those of higher
mass. This scenario was investigated in \cite{jam06} to explain the high
fraction of SNII exploding away from the on-going SF of host galaxies.\\
Using these arguments as a base, we now further discuss the progenitor mass constraints and
sequences outlined above.

\subsection{Progenitor mass sequences}
\label{seq}
Fig. 2 shows a striking sequence of increasing progenitor mass. This starts with
the SNIa arising from the lowest mass progenitors, through the
SNII, the SNIb and finally the SNIc arising from the highest mass stars. With respect
to the SNIa this is to be
expected as these events are considered to arise from WD systems, i.e. those
produced by low mass stars. Indeed this has been observed previously,
through the presence of SNIa, and absence of CC SNe within old elliptical
galaxies (e.g. \citealt{van05}).
The next group in this sequence are the
SNII. (Given the dominance of SNIIP we consider this discussion relevant for the
overall SNII population but also of that of the SNIIP.) These SNe, while
showing the expected increase of association and hence increase in progenitor mass compared to
the SNIa, are not seen to  trace bright HII regions within
hosts. We explain this result with the conclusion that these events arise from
the lower end of the mass range of CC SNe. 
This is strengthened by the fact that
these events accurately trace the recent SF. \\
Next in the sequence we find
the SNIb. These SNe only show a slightly higher correlation than the SNII, but
the difference between the SNIb and SNIc is significant. Therefore there is a
\textit{strong suggestion that SNIb arise from less massive progenitor stars
than SNIc}. 
The final transients in this sequence are the SNIc. These events are
consistent with being drawn randomly from the spatial distribution of
on-going SF, and hence are consistent with being produced by stars at the
higher end of the CC mass range.\\
While we believe that this is the
first time observationally these mass differences between `normal' CC
SNIIP and other types has been convincingly shown, these differences have been previously
predicted. Single star models
\citep{heg03,eld04,geo09} predict that CC events which have lost part of
their outer envelopes arise from more massive progenitors than those of
SNIIP. These models also predict the overall progenitor mass sequence we have
outlined, with SNIb arising from more massive stars than SNIIP, and SNIc
arising from even higher mass progenitors. While these mass differences are
likely to be more pronounced in single star scenarios,
binary system models \citep{pod92,nom96} also seem to predict some correlation of SN type with
progenitor mass. Observationally there have
been suggestions of mass differences through previous work on environments
(e.g. \citealt{van99,kel08}; AJ08), comparison of these environment studies
with galaxy models \citep{ras08}, and also hints from direct detection studies
which will be discussed below. However, the study of a large number of SNe and
their host environments we present here, together with the use of \ha\ imaging
tracing only the sites of the most massive stars, significantly strengthen this progenitor
mass sequence picture.

\subsection{The SNIIn and `impostors'}
\label{disssub}
Probably the most surprising results to emerge from this study (although this
was already shown in AJ08) is the low
correlation of SN `impostors' and SNIIn with host galaxy SF. This is shown in
Fig. 3 and the sub-type progenitor mass sequence outlined in section 4.1.
The general
consensus is that both of these transients have LBV
progenitor stars. This is because of the high mass loss rates needed to
provide the CSM required to produce the observed signs of interaction. 
It has been claimed (see e.g. \citealt{smi08}) that only the most massive
stars going through eruptive stages of mass-loss can provide this material. 
Also, a number of authors have linked LBVs to individual SNIIn and
SN `impostors' (e.g. \citealt{tru08,smi10,kie12}), while the one direct
detection of a SNIIn that exists points to a very massive star
\citep{gal07}. However, the validity of associating the majority of these transients
with LBV progenitors has been questioned by \cite{dwa11} and
\cite{koc12}. Indeed, it appears that the SNIIn and `impostor' family is extremely
heterogenous with the possibility of a significant fraction arising from lower
mass stars in dusty environments (e.g. \citealt{tho09_2,koc12}). Unfortunately our
sample sizes of these events are too small to further separate them. 
For \textit{both} SNIIn and SN `impostors' we find that they
do not trace HII regions within their galaxies and this implies that the
\textit{majority} of these events arise from relatively low mass progenitors. 
This result would seem to be inconsistent with
the majority of these transients arising from very high mass stars.
We note that this does not exclude a small fraction events such as
SN2005gl \citep{gal07}, arising from very massive stars.\\
While SNIIn are generally considered to be part of the CC SN family, there
have been several claims in the literature of SNIa-like events showing strong
signs of interaction with a dense CSM. For two SNe; 2002ic \citep{ham03b} and
2005gl 
\citep{ald06} 
it has been argued that they are part of a hybrid `Ia-IIn'
classification. While it seems unlikely that all SNIIn arise from
thermonuclear explosions, the possibility remains that there are further cases
of SNIa events which are linked to SNIIn, but their spectral features are
hidden beneath the continuum dominated spectra observed for interaction driven
SNIIn. 
This
discussion is driven by our results that SNIIn show a lesser degree of
association to SF than SNIIP, with the distribution lying 
between those of SNII and SNIa as shown in Fig. 12. 
This result could be naturally explained if a significant fraction of SNIIn
had progenitors consistent with SNIa events; i.e. lower mass than those which
explode through CC. While further examples of this hybrid class are needed to 
strengthen these arguments, we find the distributions as shown in Fig. 12
intriguing; those of SNIIn and SNIa are very similar, and statistically are
consistent with being drawn from the same parent population (KS probability $>$10\%).\\
\begin{figure}
\includegraphics[width=8.5cm]{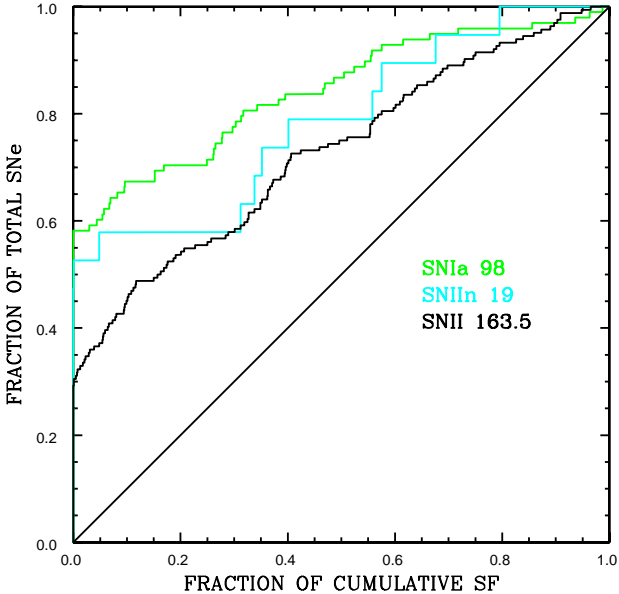}
\caption{Cumulative pixel statistics with respect to \ha\ host galaxy emission
for the SNIa (97 events) shown in green, the SNIIn (19) in cyan, and the SNII
(163.5) shown in black. The SNIIn distribution falls almost exactly in the
middle of the SNIa and SNII populations.}
\end{figure}
If one wishes to retain high mass LBV stars as the progenitors of both the `impostors'
and SNIIn then one needs to find reasons why some high mass stars explode
far from HII regions. Indeed, \cite{smi11_2} specifically
predict that SNIIn, given the assumption of LBV progenitors, will be found to
occur closer to HII regions than SNIIP. One could hypothesize that we do not observe these transients 
near to bright HII because the gas geometry is such that the transients are
observed but the tracers we use of SF are not. It follows that one would, in
addition, observe these events explode outside near-UV traced SF regions. However, earlier
we showed that they follow near-UV emission of their host galaxies better than
\ha. If this were an issue of dust extinction we may expect it to be
\textit{worse} at UV wavelengths (than in the optical where we detect
\ha). This closer association to near-UV emission appears to strengthen the
conclusion that these events are arising from lower mass progenitors.\\ 
For the
SN `impostors' there is one strong caveat to these conclusions. Events
classified as `impostors'
are generally at the low end of the CC luminosity function. Hence, there will
be a strong selection effect against finding these transients within bright
background regions (this is seen in the median recession velocity of these transients
within our sample as listed in table 6). However, we note that if this
were the case, and the `impostors' \textit{intrinsically} followed
distributions of high mass stars, this would imply
that even in very nearby galaxies we are missing the vast majority of these
events, and therefore that the SN `impostor' rate is considerably higher than
assumed.
For the SNIIn this effect does not appear a viable option as these
events are often \textit{brighter} than `normal' SNII.

\subsection{Comparison with Smith et al. (2011)}
\label{comp}
\cite{smi11_2} recently used local SN rates from the Lick Observatory
Supernova Search \citep{li11} to investigate mass ranges for different SN
types by comparison to a standard IMF, and attempted to constrain whether
single stars, binary systems or a hybrid of both could best explain nearby SN
rates. These authors
concluded that single stars cannot account for the relative rates of different
CC SNe. They tested various scenarios and 
suggested mass constraints for different SN types. In their favoured
scheme (a hybrid of single and binary systems, their Fig. 7) SNIIP, SNIIL
and SNIIn are produced by single stars and form a sequence of increasing
progenitor mass. SNIIb and SNIb are produced by binaries and form over the
entire CC mass range. Finally SNIc are produced by both single and binary
systems and have progenitors of high mass. We now
discuss each SN type in turn with respect to our results.\\
SNIIP arise from the lowest mass bin in all scenarios of \citeauthor{smi11_2}, which
is consistent with our results. In their preferred scenario, SNIIL trace
slightly higher mass stars than SNIIP, again consistent with our results and
the mass sequence presented in section 4.1. In \textit{all} presented
scenarios these authors predict (or assume) that SNIIn arise from higher mass
progenitors than SNIIP. We see the opposite trend with the SNIIn showing 
the \textit{lowest} degree of association to
the on-going SF. 
\citeauthor{smi11_2} constrain SNIIb to arise
from binaries from the entire CC mass range. Our results suggest that these SNe
arise from higher mass stars than both the SNIIP and SNIIL, but given the
low numbers, it is hard to discriminate between different scenarios. SNIb arise
from moderately massive stars in both our and their mass sequences, while SNIc
arise from the most massive stars. We conclude that overall our results are
consistent with those of \citeauthor{smi11_2}, and their hybrid scenario of single
and binary progenitors, with one major exception; the SNIIn.\\

\subsection{Consistency with direct detections and previous results}
\label{prev}
In the introduction we outline the method of `direct detections' for
constraining SN progenitor masses. This technique affords much information on
individual events, but is restricted by small number
statistics. Here we compare those results with those currently presented. 
The only type of CC SN to be
studied in a statistical sense with relation to direct detections are the
SNIIP. \cite{sma09} used the available progenitor mass constraints (plus upper
limits) to estimate lower and upper limits for SNIIP progenitors of 8.5 and
16.5 \msun\ respectively. Here we have constrained that the majority of SNIIP
arise from the low end of the CC mass range. This is completely consistent with the
direct detections to date. 
We note that our results are completely independent
of those from \cite{sma09}, and hence strengthen the argument for SNIIP
progenitor masses extending down to low values. Recently a number of
progenitor identifications have been published for SNIIL
(\citealt{eli10,eli11}, although the clear identification of 
one as a definitive IIL has been questioned by \citealt{fra10}).
These detections appear to point towards progenitor stars above
15\msun, and also suggest yellow supergiant progenitors rather than
  the red supergiant progenitors determined for SNIIP \citep{sma09}. 
Therefore, we are seeing the same trend; SNIIL have higher mass
progenitors than SNIIP. A similar trend has been observed for the SNIIb. Two
very close explosions have occurred with IIb spectra; SN1993J and SN2011dh. In
both cases, whether one assumes a single or binary scenario, the progenitors
appear to be in the 15-20\msun\ range \citep{ald94,mau04,mau09,mau11,van11}. In terms of
differences to the SNIIP, this appears consistent with our results; SNIIb
arise from more massive stars than SNIIP (whether binary or single
systems). There are neither enough direct detections, nor SNe within our
samples to evaluate differences between the SNIIL and SNIIb.\\
There has been one definitive direct detection of a SNIIn; SN2005gl
\citep{gal07}. This constrained the progenitor star to be extremely massive;
$\sim$50\msun. If we were to generalise to the whole SNIIn population and
conclude that all SNIIn have similarly massive
progenitors this result would seem very inconsistent with our environment
constraints. However, SNIIn are extremely heterogenous and different events
may have different origins; the evidence of
interaction simply betrays the close (circum-stellar) environment into which an
event explodes. Hence, it is probable that there exists multiple channels
through which transients display SNIIn-type spectra. We therefore conclude
that SN2005gl must be a special case\footnote{While there have been a number of
very luminous SNIIn reported in the literature, some appearing to require very
massive progenitors, these are likely to be overrepresented. 
This is because, a) their high luminosity simply makes them easier
to detect, and b) due to their brightness they are more likely to be studied
in detail and hence progenitor constraints are made.}, and that the majority of nearby SNIIn have
lower mass progenitors, as suggested by our statistics.\\
Unfortunately we are still awaiting a direct detection of
a SNIbc \citep{sma09b}. This lack of detection (there have been a number
of upper limits) has been argued as evidence that SNIbc arise from lower mass
binary systems and not single stars. However, the low number statistics, plus
the uncertain nature of current modelling of WR stars (the possible 
progenitors of SNIbc; e.g. \citealt{gas86}) means that constraints are less
reliable than for SNIIP.
With a lack of information from direct detections, our work, together with
that presented in \cite{kel08}, appears to be the only convincing evidence
that SNIbc arise from more massive progenitors than SNII.\\
Besides our own previous work on environments,
there have been a number of investigations on the positions of SNe
within their host galaxies. Of particular relevance is the work published by
\cite{kel08} and \cite{kel11}. The first of these works employed a very similar
pixel technique to that used by ourselves but using host galaxy $g'$-band
light in place of \ha\ emission as a tracer of massive stars. They found
similar results; the SNIc much more frequently explode
on bright $g'$-band regions of their hosts, less so the SNIb and even less so the SNII.
The similarity of these results is encouraging.
\cite{kel11} also looked at colours of environments, finding that the host
environments of SNIb and SNIc are intrinsically bluer than those of SNII and
use this to argue further for higher mass progenitors, again consistent with
the current and previous work using pixel statistics. Finally, these authors
found, as we do, that the environments of SNIIn are similar to those of the overall
SNII population.

\subsection{Progenitor metallicity}
\label{progZ}

\begin{figure}
\includegraphics[width=8.5cm]{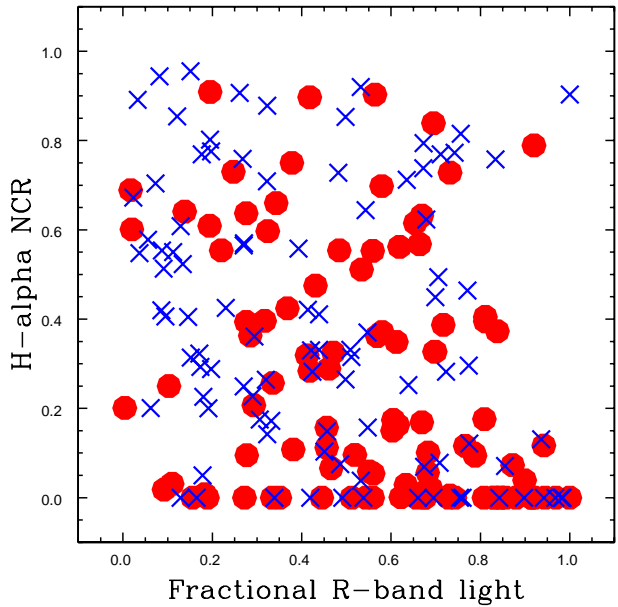}
\caption{\ha\ NCR pixel values plotted against 
a normalised parameter tracing the radial location of each SN, with 0 
meaning a central location, and 1 meaning an outlying location.
The SNIIP plus SNII (all those with no further
type classification) are shown as red circles, while the SNIbc are plotted 
as blue crosses.}
\end{figure}

Thus far we have not outlined the role of metallicity within our discussion of the 
implications and conclusions gained from our results. Increased progenitor metallicity is thought 
to affect the evolution of stars through increasing the metallicity dependent line driven winds
and hence increasing the rate of mass loss (e.g.  \citealt{pul96,kud00,mok07}). An increase in 
metallicity could thus increase the ability of a star to lose sufficient amounts of its outer
envelope and explode as a SNIbc rather than SNII (see e.g. \citealt{heg03,eld04,geo09}, for predictions
on how the ratio of SN types changes with metallicity). Hence, the effect 
environment metallicity has on our NCR statistic may need to be taken into account. 
It has been found by many authors that SNIbc are
more centrally concentrated within host galaxies than SNII (e.g. \citealt{bar92,van97,tsv04,hak08,boi09,and09}). 
Until recently these radial differences have been almost exclusively ascribed to metallicity effects, 
with the central parts of galaxies being more metal rich than the outer environments (see e.g. \citealt{hen99}).
Therefore one may speculate that higher NCR values for SNIbc (with respect to SNII) are
due to the centralisation of these events where more central SNe fall on
brighter HII regions. Hence, the hypothesis would be that the relatively higher pixel values for SNIbc 
is actually a metallicity and not a mass effect. We rule this out on the basis of the following
arguments:\\
a) The validity of the assumption that the brightest regions of SF within galaxies are centralised is
unclear. We see within our sample a huge variety of distributions of SF and this 
is also shown in the 3 example images displayed in Fig. 1. We can see a clear case (that on the right)
of a galaxy where the dominant \ha\ regions are in the spiral arms and not the central parts of the galaxy.\\
b) \cite{hab10} (also see Habergham et al. in preparation; these studies used the same sample
we present here) showed that the centralisation of SNIbc is dominated 
by those SNe occurring within disturbed galaxies, where one observes shallower if any metallicity gradients \citep{kew10}.
Therefore, even if there were some correlation between radial position and NCR value, this could
not be explained by metallicity.\\
c) In \cite{and10}, using host HII region spectroscopy of a random subset of the SNe we use for the current sample, we
showed that there is no significant metallicity difference between the 
environments (and hence progenitors) between SNIbc and SNII.\\
To test this hypothesis further we show in Fig. 13 a scatter plot of the \ha\ NCR values for SNII and SNIbc 
against the radial positions of each SN with respect to host galaxy $R$-band light (a full analysis of these
radial distributions will be presented in Habergham et al. in preparation). If this hypothesis of higher NCR values
being centralised were correct, for both samples we would expect to see a clear trend of more events in both the
top left and bottom right corners of the plot. While there appears to be some evidence that those SNe having
NCR values of 0 are more often found in the outer regions of their hosts (the bottom right part of Fig. 13) this
does not seem to be strong, and we see no evidence for the highest pixel counts 
arising from the most central regions.\\
We conclude, through the arguments outlined above plus the distributions shown in Fig. 13, that metallicity
cannot explain our results of an increased association of SNIbc to SF with respect to SNII.

\subsection{Progenitor binarity}
\label{progbin}
A major debate within the SN community currently hinges on whether 
SNe arising from  progenitor stars that have had their envelopes stripped before explosion
arise from single or binary star systems. In the single star scenario (see
\citealt{heg03,eld04,geo09} for such models), stars more massive than
25-30\msun\ lose their envelopes through strong stellar winds, the strengths
of which
are dependent on metallicity. In the binary scenario (see e.g. \citealt{pod92,nom96})
progenitor masses for the SNIbc (plus SNIIb and possibly SNIIL) extend down
further in mass as envelope removal is aided by binary companions. The lack
of direct detections of SNIbc would seem to favour binary progenitors, but as
we discuss above, thus far this is inconclusive. For at least one CC
SN; 1993J, convincing evidence for a binary progenitor system has been
observed \citep{mau09}.\\ 
We conclude that overall the
SNIbc population has higher mass progenitor stars than SNII. However, as
noted, this does not differentiate between binary or single
scenarios. If one envisages an IMF from 8 to 100\msun\ with single stars up to
e.g. 25\msun\ (generally the highest mass red giant stars; \citealt{lev07})
producing SNIIP (and possibly other SNII), while stars in the full
mass range that are binaries explode as SNIbc, then on average progenitors of
SNIbc would still be more massive than their SNII counterparts. We
speculate here that our results are consistent with SNIb arising from
binaries or single stars in an intermediate mass range (above the SNIIP but below the SNIc), 
with the possibility that SNIc could arise from more massive
binaries, very massive single stars, or a combination of both.\\
A recent study by \cite{eld11} specifically looked at the production of CC SNe
through binaries and how the initially most massive stars explode and `kick'
their lower mass companions with velocities meaning that the lower mass stars
explode further away from their places of birth than their high mass
companions. This is then broadly consistent with the results we find in terms
of the increasing association to SF of II to Ib and finally Ic. However, we
note that these models only predicted offsets of at most 100 pc from their
birth place, distances that are generally not probed by the present study.\\
A final question that can be asked in this section is whether a change in the
percentage of close binaries with environment could explain our results 
without any need for progenitor
mass differences. One may imagine that in
denser environments; e.g. bright HII regions, the fraction of close binaries
may differ from that in the field. However, 
we know of no theoretical models or observational constraints which
shed light on whether this explanation is feasible, or even on the sense 
of any such effects.
It is possible that the close binary fraction
would be higher in denser regions solely because of the higher number of stars
in close proximity during SF, but equally, a dense
environment may prevent the formation of close binaries. \textit{If} 
the fraction of close binaries were higher within HII regions then
this could explain our result without resorting to progenitor mass
differences. However, we currently see no reason to favour this
interpretation over the most logical assumption that an increasing association
to HII regions means higher mass progenitor stars.

\section{Conclusions}
\label{conc}
We have presented constraints on the relative progenitor ages, and therefore
implied masses of the different CC SN types, using a pixel statistics 
method applied to \ha\ and near-UV imaging of their host galaxies. 
Using the assumption
that a higher degree of association of a population to host on-going SF means shorter
lifetimes
we constrained differences in
the progenitor masses of CC SNe. Where distributions show a low degree of
association to the line emission, we also investigated the association to a
longer lived SF component; the recent SF, as traced by near-UV emission. We
now list our main results and conclusions achieved with this study.

\begin{itemize}
\item SNIbc show a much higher degree
of association to the on-going SF of their host galaxies than the SNII. Whether the main
progenitor channel for the former is a single or binary scenario, this implies
for both channels that \textit{SNIbc arise from shorter lived and therefore more
massive progenitor stars than SNII.}
\item Including the SNIa for comparison, we find a strong trend of
increasing progenitor mass, starting with the SNIa from the lowest, through
the SNII, then the SNIb, with the SNIc arising from the highest mass progenitor stars.  
\item The SNIc show the highest degree of association of all SNe types, arguing
that these SNe
arise from the most massive stars that explode as visible transients. Indeed, they are seen
to correlate with SF on the shortest timescales much more than the SNIb. This argues that SNIc 
arise from more massive stars than SNIb.
\item While the SNIIP \textit{do not} follow the underlying distribution of
the on-going host galaxy SF, these SNe do follow the distribution  of
recent SF as traced by near-UV emission. This result suggests that the
majority of the progenitors of these SNe populate the lower mass range of
massive stars that explode as CC SNe.
\item We find that the SNIIn show an even lesser degree of association with
the on-going SF than the SNIIP. The most logical explanation for
this result is that the majority of these SNe \textit{do not arise from the deaths of the most massive
stars, in contradiction to several previous studies of these SNe.}
\end{itemize}

\section*{Acknowledgments}
The referee Justyn Maund is thanked for constructive comments.
Paul Crowther, Francisco Forster, Santiago Gonzalez and Maryam Modjaz 
are thanked for useful discussion.
J. P. Anderson \&\ M. Hamuy acknowledge support by CONICYT through
FONDECYT grant 3110142, and by the Millennium Center for
Supernova Science (P10-064-F), with input from `Fondo de
Innovaci\'on para la Competitividad, del Ministerio de
Econom\'ia, Fomento y Turismo de Chile'. 
P. A. James \&\ S. M. Habergham acknowledge the UK Science and 
Technology Facilities Council for research grant, and research 
studentship support, respectively.
This research
has made use of the NASA/IPAC Extragalactic Database (NED) which is operated
by the Jet 
Propulsion Laboratory, California
Institute of Technology, under contract with the National Aeronautics and
Space Administration and of data provided by the Central Bureau for Astronomical Telegrams. 

\bibliographystyle{mn2e}

\bibliography{Reference}

\appendix

\section[]{SN and host galaxy data}

\begin{table*}
\begin{tabular}[t]{cccccccc}
\hline
SN & Host galaxy & Galaxy type &V$_\textit{r}$ (\kms ) & SN type & NCR value& Telescope & Reference\\
\hline				       
1936A  & NGC 4273  & SBc & 2378 & IIP & 0.362 & LT & \\
1982R  & NGC 1187  & SBc & 1390 & Ib  & 0.000 & LT & \\
1983V  & NGC 1365  & SBb & 1636 & Ic  & 0.330 & ESO& \\
1984L  & NGC 991   & SABc& 1532 & Ib  & 0.853 & LT & \\
1984F  & MCG +08-15-47 & Im &  2254 & II &  0.112 & JKT & SAI SN catalogue\\
1984I  & ESO 323-G99&SABc& 3219 & Ib  & 0.000 & ESO& \\
1985P  & NGC 1433  & SBab& 1075 & IIP & 0.000 & ESO & \\
1987B  & NGC 5850  & SBb & 2556 & IIn &0.000 & ESO & \\
1988H  & NGC 5878  & SAb & 1991 & IIP & 0.000 & LT & \\
1990B  & NGC 4568  & SAbc& 2255 & Ic  & 0.776 & JKT & \\
1990I  & NGC 4650A &     & 2880 & Ib  & 0.000 & ESO & \\
1990W  & NGC 6221  & SBbc& 1499 & Ic& 0.642 & ESO & \cite{van99}\\
1993N  & UGC 5695  & S  & 2940 & IIn & 0.000 & ESO & \\
1994N  & UGC 5695  & S  & 2940 & II  & 0.000 & ESO & \\
1994W  & NGC 4041  & SAbc& 1234 & IIn & 0.795 & LT & \cite{fil94}\\
1994ai & NGC 908   & SAc & 1509 & Ic  & 0.523 & ESO & \\
1995X  & UGC 12160 & Scd & 1555 & II  & 0.903 & LT &  \\
1996D  & NGC 1614  & SBc & 4778 & Ic  & 0.361 & JKT &  \\
1996N  & NGC 1398  & SBab& 1396 & Ib  & 0.315 & ESO & \\
1996cr & ESO 97-G13& SAb & 434  & IIn & 0.575 & ESO & \\
1997B  & IC 438    & SAc & 3124 & Ic  & 0.644 & ESO & \\
1997D  & NGC 1536  & SBc & 1217 & IIP &0.000 & ESO & \cite{elm03}\\
1999bg & IC 758    & SBcd& 1275 & IIP & 0.632 & LT & \\
1999el & NGC 6951  & SABbc&1424 & IIn & 0.048 & LT & \\
1999ga & NGC 2442  & SABbc&1466 & IIL & 0.349 & ESO & \\
1999go & NGC 1376  & SAcd& 4153 & II  & 0.363 & ESO &\\
2000H  & IC 454    & SBab& 3945 & Ib & 0.623 & ESO & \cite{bra02}\\
2000P & NGC 4965   & SABd& 2261 & IIn & 0.000 & ESO &\\
2000cl & NGC 3318  & SABb& 2775 & IIn & 0.312 & ESO &\\
2000fn & NGC 2526  & S   & 4603 & Ib &  0.282 & ESO &\\ 
2001X  & NGC 5921  & SBbc& 1480 & IIP & 0.698 & LT & \\
2001bb & IC 4319   & SAbc& 4653 & Ic  & 0.282 & ESO &  \\
2001db & NGC 3256  & Pec & 2804 & II  & 0.319 & ESO &\\
2001du & NGC 1365  & SBb & 1636 & IIP  & 0.101 & ESO & \\
2002J  & NGC 3464  & SABc& 3736 & Ic  & 0.331 & ESO & \\
2002ao & UGC 9299  & SABd& 1539 & Ib & 0.000 & LT & \cite{pas08}\\
2002hc & NGC 2559  & SBbc& 1559 & IIL& 0.948 & ESO & \\
2002hy & NGC 3464  & SABc& 3736 & Ib  & 0.411 & ESO &\\
2002jp & NGC 3313  & SBb & 3706 & Ic  & 0.903 & ESO &\\
2003I  & IC 2481   & S  & 5322 & Ib  & 0.000 & ESO & \\
2003dv & UGC 9638  & Im  & 2271 & IIn & 0.000 & LT & \\
2003gm & NGC 5334  & SBc & 1386 & impostor&0.000& LT & \cite{pat03}\\
2003id & NGC 895   & SAcd& 2288 & Ic  & 0.252 & LT &  \\
2003lo & NGC 1376  & SAcd& 4153 & IIn & 0.000 & ESO &\\
2005bf & MCG +00-27-05& SBb& 5670& Ib & 0.069 & ESO &\\
2004cc & NGC 4568  & SAbc& 2255 & Ic & 0.608 & JKT & \\
2004ch& NGC 5612   & SABb& 2699 & II  & 0.371 & ESO &\\
2004cm & NGC 5486  & SAm & 1390 & IIP & 0.201 & JKT & \\
2004ds & NGC 808   & SABbc&4964 & IIP & 0.250 & ESO &\\
2004ez & NGC 3430  & SABc& 1586 & IIP & 0.094 & LT & \\
2004gn & NGC 4527  & SABbc & 1736 &Ic & 0.558 & INT &\\
2005U  & ARP 299  & IBm/SBm& 3088& IIb & 0.672 & INT &\\
2005bq & IC 4367  &  SBbc& 4090 & Ic  & 0.227 & ESO & \\
2005dg & ESO 420-G03 & SAbc & 4131 & Ic & 0.249 & ESO &\\
2005my & ESO 302-G57& SABc&4441 & II  & 0.839 & ESO &  \\
2006T  & NGC 3054  & SABbc&2426 & IIb & 0.000 & LT &  \\
2006ba & NGC 2980  & SABc &5720 & IIb & 0.227 & ESO &\\
2006bc & NGC 2397  & SABb& 1363 & IIP  & 0.750 & ESO &\\
2006bv & UGC 7848  & SABcd&2513 & impostor & 0.000 & LT & \cite{smi11}\\
2006fp & UGC 12182 & S & 1490 & impostor & 0.965 & LT & \cite{blo06}\\
2006my & NGC 4651  & SAc & 788& IIP& 0.553 & LT & \\
2006lv & UGC 6517  & Sbc & 2491& Ib/c & 0.000 & JKT &\\
2007C  & NGC 4981  & SABbc&1680 & Ib  & 0.264 & LT & \\
2007Y  & NGC 1187  & SBc  & 1390 & Ib & 0.000 & LT & \\

\hline
\end{tabular}
\end{table*}

\setcounter{table}{0}

\begin{table*}
\begin{tabular}[t]{cccccccccc}
\hline
SN & Host galaxy & Galaxy type &V$_\textit{r}$ (\kms ) & SN type & NCR value& Telescope & Reference\\

\hline
2007aa & NGC 4030  & SAbc& 1465& IIP  & 0.117 & LT & \\
2007ay & UGC 4310  & SAm & 4355 & IIb & 0.000 & ESO &  \\
2007fo & NGC 7714  & SBb & 2798& Ib & 0.149 & LT &  \\
2007gr & NGC 1058  & SAc & 518 & Ic & 0.157 & JKT &\\
2007od & UGC 12846 & Sm& 1734& IIP & 0.000 & LT &  \\
2007rw & UGC 7798  & IBm & 2568 & IIb & 0.784 & ESO & \\
2008M  & ESO 121-G26  & SBc& 2267 & IIP & 0.789 & ESO &  \\
2008S  & NGC 6946  & SABcd& 40 & IIn &0.000 &LT & \cite{smi11}\\
2008V  & NGC 1591  & SBab &4113 & IIb & 0.685 & ESO &  \\
2008W  & MCG -03-22-07& Sc& 5757 & IIP & 0.005 & ESO &\\
2008X  & NGC 4141  & SBcd& 1897& IIP & 0.609 & LT &  \\
2008aq & MCG -02-33-20& SBm &2390 & IIb & 0.909 & ESO & \\
2008bo & NGC 6643  & SAc  & 1484 & IIb  & 0.103 & LT &  \cite{nav08} \\
2008cn & NGC 4603  & SAbc & 2592 & IIP & 0.008&ESO& \cite{eli09}\\
2008dv & NGC 1343  & SABb & 2215 & Ic &  0.802 & LT & \\
2008dw & UGC 8932  & Im & 3728   & II & 0.826 & JKT & \\
2008ij & NGC 6643 & SAc & 1484   & IIP & 0.096 & JKT & \cite{cha08}\\
2008im & UGC 2906  & Sb   & 2494 & Ib  & 0.000 & LT &  \\
2008iz & NGC 3034  & I0   & 203  & II & 0.496  & INT &\\
2009dd & NGC 4088  & SABbc& 757  & II & 0.601  & JKT &\\
2009kr & NGC 1832  & SBbc & 1939 & IIL & 0.890 & INT &\\
2009hd & NGC 3627  & SABb & 727  & IIL  & 0.398 & INT & \cite{eli11}\\
2009jf & NGC 7479  & SBc  & 2381 & Ib  & 0.464 & JKT &\\
2009js & NGC 918   & SABc & 1507 & IIP & 0.000 & JKT &\\
2010O  & ARP 299   & IBm/SBm&3088& Ib  & 0.421 & INT &\\
2010P  & ARP 299   & IBm/SBm&3088& Ib/IIb*&0.406& INT &\\
2010br & NGC 4051  & SABbc& 700 & Ib/c & 0.000 & INT & \\
2010dn & NGC 3184  & SABcd& 592 & impostor & 0.000 & INT& \cite{smi11}\\
2011aq & NGC 1056  & Sa   & 1545 & II &  0.554 & JKT & \\
2011ca & NGC 4495  & Sab  & 4550 & Ic & 0.121  & INT & \\
2011dh & NGC 5194  & SAbc & 463  & IIb & 0.000 & INT & \\
2011fd & NGC 2273B & SBcd & 2101 & IIP & 0.000 & JKT & \\
\hline
\end{tabular}
\caption{SN and host galaxy information for all those `new' SNe analysed in
this work (the `old' events being those presented in AJ08). In column 1 we
list the SN name followed by its host galaxy in column 2. Then the host Hubble
type is given in column 3 and its recession velocity in column 4. The SN type
is then listed followed by the \ha\ NCR value. In column 7 we show the
telescope used to obtain the observations for that particular host galaxy.
Cases where a different SN type was used from that
documented on the Asiago catalogue are listed together with a reference in
column 8. 
Note, SN1987B is listed as `IInL' in the Asiago catalogue. Here we 
put this in the `IIn' class for our investigation.
Finally, we note that in AJ08
the NCR values were wrongly listed as 0.000 for SN1964L, SN1986I and SN1999bu. The
true values are 0.453, 0.005 and 0.226 respectively.}
\end{table*}

\label{lastpage}

\end{document}